\documentclass[a4paper,11pt]{article}
\pdfoutput=1 

\usepackage{jcappub} 

\usepackage[toc,page]{appendix}
\usepackage{graphics}
\usepackage{xcolor,cancel}
\usepackage{caption}
\usepackage{subcaption}
\usepackage{rotating}
\usepackage{pstricks}
\usepackage{color}
\usepackage{amsfonts}
\usepackage{mathrsfs}
\usepackage{epsfig}
\usepackage{amsmath,amssymb,amsthm,graphicx,latexsym}

\definecolor{ao}{rgb}{0.0, 0.5, 0.0}
\usepackage[normalem]{ulem}
\newcommand{\stkout}[1]{\ifmmode\text{\sout{\ensuremath{#1}}}\else\sout{#1}\fi}
\definecolor{dblue}{rgb}{0,0,0.6}
\definecolor{dred}{rgb}{1,0.08,0.58}

\title{\textcolor{dblue}{A New Class of  Non-canonical Conformal Attractors for Multi-field Inflation}}

\author{{Tony Pinhero}}
\author{{and Supratik Pal}}

\affiliation{\textit{Physics and Applied Mathematics Unit, Indian Statistical Institute, \\203 B.T. Road, Kolkata 700 108, India}}

\emailAdd{tonypinheiro2222@gmail.com}
\emailAdd{supratik@isical.ac.in}

\abstract{We propose a new broad class of multi-field non-canonical inflationary models as an extension of multi-field conformal cosmological attractors. This also generalizes the recently discovered class of non-canonical conformal attractors for single field inflation. Kinetic terms of this class of models are phenomenologically arising from ${\cal N}=1$ supergravity  and from ${\cal N}=1$ superconformal theory, with two conformal scalar compensator fields in the latter. We show that the inflationary dynamics and predictions of this class of models are stable with respect to the significant modification of both radial and angular part of the potential, but it is very sensitive to its minuscule modification in the geometry of the field space metric. We also show that our framework can pass the latest observational constraints set by Planck 2018.}

\begin{document}
\maketitle
\flushbottom
\section{Introduction}\label{sec_intro}

Cosmological attractor scenario for the  inflationary models are being thoroughly developed in the past couple of years \cite{kallosh2013universality, kallosh2013multi,kallosh_nonminimal_attra2013,kallosh2013sup_alpha_attra,unity_of_cosmo_attracts,escher_in_sky,hyperbolic_geometry_of_attrctors,seven_disc_manifold,pinhero2017non-cano-con-attra-sing-inf,b_mode,ana_achucarro_multi-field_alph_attra,hypernatural_inflation,scalisi_alpha_scale,scalisi_desitter_landscape,pole_nflation,yakrami_dark_energy}. In conformal multi-field attractor formulation \cite{kallosh2013multi}, the study is mainly based on two scalar fields: a canonically normalized radial field $\psi$ and a non-canonically normalized angular field $\theta$ with its kinetic term taking the form $3\sinh^{2}\left(\psi/\sqrt{6}\right)(\partial_{\mu}\theta)^{2}$. The function in front of $\partial_{\mu}\theta$ in the corresponding Lagrangian grows exponentially for large $\psi$ and during inflation, speed of motion for both the fields $\psi$ and $\theta$ are effectively suppressed by the same factor. Since the range of evolution for the angular field $\theta$ is very small (of the order of one) compared to the very large range of evolution of the radial field $\psi$, this angular field $\theta$ rapidly rolls down to the minimum of the potential (valley) with respect to the $\theta$ field. Then the subsequent evolution effectively becomes a single field evolution and the radial field serves the role of inflaton. As a result, in such a two-field system, one can observe that, inflaton first rolls down from the ridge to the valley and then slowly rolls down to the minimum of the potential along the valley.  

A similar kind of dynamics can be found in the multi-field Higgs inflation with non minimal coupling $\xi>0$ \cite{kaiser_mult_field_higgs, greenwood_mult_higgs_infla} with a different mechanism. Here the function in front of $\partial_{\mu}\theta$ does not grow exponentially for large $\psi$, but it grows very slowly and reach the constant $\xi^{-1}$. As a result mass of the angular field becomes large at large $\psi$ and it rapidly rolls down to the minimum of its potential, and the rest of the evolution is dominated by the radial field $\psi$ which can be consider as the inflaton.

In the two types of models mentioned above, even though these models are multi-field in its construction, one can consider this as an effectively single field evolution since the later stage of evolution is only dominated by the single radial field and also one can readily neglect the multi-field effects during the calculations of inflationary parameters and can conclude that its predictions are that of the single field inflation. In this paper we are going to demonstrate, somewhat a different mechanism and a different dynamics for the multi-field inflation by generalizing the idea of conformal multi-field case. This paper is a continuation of our previous work \cite{pinhero2017non-cano-con-attra-sing-inf}, where we studied the possibility of having a set of non-canonical kinetic terms, for the single field, participating in the construction of the so-called “conformal inflationary attractors” \cite{kallosh2013universality}. There we where  mainly motivated by the Planck 2015 and Planck 2018 data \cite{planck2018_inflation, planck2015_inflation, planck2015_non-Gaussianity}, which confirmed the  some scale dependence in the power spectrum at 5-$\sigma$,  so, one can in general generate scale-dependent power spectrum from non-canonical models, which have become more relevant than ever.  Here, we generalize this idea of \cite{pinhero2017non-cano-con-attra-sing-inf} to multi-field sectors in the same way. From a purely theoretical point of view, this generalization is necessary, since the supersymmetric theories are nourished with a lot of scalar fields in its construction, and their kinetic terms are non-canonical in its generalized form. On the other hand, from the observational point of view, Planck data \cite{planck2018_inflation, planck2015_inflation, planck2015_non-Gaussianity} puts very tight constraints on the multi-field inflationary models. For example, usually, multi-field inflationary models predicts large local non-Gaussianity, which is disfavored by the recent Planck data \cite{planck2018_inflation, planck2015_inflation, planck2015_non-Gaussianity}. So in this paper, we aim to construct a new class of multi-field inflationary models, which bypass all those tight constraints provided by the recent Planck results. We accomplish this task by generalizing the well-known multi-field model named multi-field conformal attractors \cite{kallosh2013multi} by adding non-canonical non-homogeneous kinetic terms into it. 

In our model, both radial field $\psi$ and angular field $\theta$ are non-canonically normalized, and the functions in front of the terms $(\partial_{\mu}\psi)^{2}$ and $(\partial_{\mu}\theta)^{2}$ are non-homogeneous. Due to the behavior of  this non-canonical non-homogeneous kinetic terms, the speed of the angular field is hugely suppressed for the large value of $\psi$. As a result, instead of rolling to the minimum of the valley of the potential, one can see that a rolling on the ridge of the potential occurs due to the combined evolution of these fields. Thus, multi-field effects come into play in this class of models intrinsically and the predictions  obviously are that of the multi-field inflation. One may see that this dynamics i.e., rolling on the ridge, is somewhat similar to that of the dynamics studied in the multi-field $\alpha$-attractor case \cite{ana_achucarro_multi-field_alph_attra}. The basic difference between multi-field $\alpha$-attractor model and our model is arising from in its supergravity construction. In multi-field $\alpha$-attractor scenario both radial (canonically normalized) and angular fields (non-canonically normalized) appear as pairs, i.e., they are originated from the same superfield, just like models studied with axion-dilaton pairs \cite{axion-dilaton_starobinsky,axion-dilaton_brandenberger}. But in our case these fields are arising from two different superfields of the K\"ahler potential and their partners are stabilized during inflation.

This paper is organized as follows. In section \ref{section_base_set}, we give a phenomenological construction of Lagrangian which contains non-canonical kinetic terms from $\mathcal{N}=1$ supergravity and from $\mathcal{N}=1$ superconformal theory, in terms of original conformal variables. In section \ref{Non-canonical multi-field inflationary attractors}, we discuss the phenomenology of non-canonical multi-field conformal attractors and in section \ref{inflationary_dynamics}, we study the dynamics of the fields during inflation, followed by the inflationary parameter calculations and their predictions in section \ref{inflationary_parameters}. Finally section \ref{summary_and_conclusions} is dedicated for summary and conclusions.

\section{Basic setup}\label{section_base_set}
 
 The aim of this section is to show that, the form of the non-standard kinetic terms, that we intend to consider in the participation of the construction of multi-field conformal attractors, have a phenomenological  supergravity origin and to investigate the possibility of embedding such a Lagrangian in a superconformal theory. In order to construct a supergravity model of inflation with non-homogeneous non-canonical kinetic term, we need to consider some sort of new shift symmetry in the K\"ahler potential to suppress the exponential growth of the potential due to the prefactor $\sim e^{K}$ in the scalar supergravity potential. The traditional form of the shift symmetry proposed in the context of chaotic inflation was $\Phi\rightarrow\Phi+C$ for the inflaton superfield $\Phi$ and the K\"ahler potential which obeys this shift symmetry has the form $K=\frac{1}{2}\left(\Phi-\Phi^{*}\right)^{2}$ and it will end up with the canonical kinetic term for the inflaton field \cite{kawasaki2000naturalchaotic}. In order to get a non-canonical kinetic term in the Lagrangian, one can generalize this shift symmetry to the form $\Phi^{n}\rightarrow\Phi^{n}+C$ and hence the corresponding  K\"ahler potential $K=c\left(\Phi^{n}-\Phi^{*n}\right)^{2}$ leads to the non-canonical kinetic term of the form $L_{\text{kin}}\sim c\phi^{2n-2}\partial_{\mu}\phi\partial^{\mu}\phi$ for the real inflaton field $\phi$ \cite{linear_running_kinetic_inflation_takahashi,running_kinetic_inflation_takahashi}. As we are interested to study the participation of a set of non-canonical non-homogeneous kinetic terms (say for example $\left(1+K^{(2)}\phi+K^{(3)}\phi^{2}+\dots\right) \partial_{\mu}\phi\partial^{\mu}\phi$ in terms of real inflaton field variable $\phi$) in the well known construction so called 'conformal attractors'  we further generalizes the above mentioned shift symmetries to the following form \cite{pinhero2017non-cano-con-attra-sing-inf}:
 \begin{equation}\label{shift_symmetry_single_field}
 \sum_{\substack{n}}^{l}K^{(n)}\Phi^{n}\rightarrow \sum_{\substack{n}}^{l}K^{(n)}\Phi^{n}+C^{l}
 \end{equation}
The K\"ahler potential which obeys this shift symmetry has the explicit form: 
\begin{equation}\label{Kahler_pot_single_field}
K=-\left[\sum_{\substack{n}}^{l}K^{(n)}
\left(\Phi^{n}-\Phi^{*n}\right)\right]^2+SS^{*}-
\zeta (SS^*)^{2}
\end{equation} 
and this will boils down to the non-canonical non-homogeneous kinetic term in the Lagrangian as shown in our previous work \cite{pinhero2017non-cano-con-attra-sing-inf}.  One can also further generalize this K\"ahler potential Eq.(\ref{Kahler_pot_single_field}) to multi-field sectors and 
 we start with  this new multi-field  K\"ahler potential  and a superpotential, which has the form 
\begin{equation}\label{eq_Kahler_potential}
K=-\sum_{\substack{i}}^{M}\left[\sum_{\substack{n}}^{l}K_{i}^{(n)}
\left(\Phi_{i}^{n}-\Phi_{i}^{*n}\right)\right]^2+SS^{*}-
\zeta (SS^*)^{2},~~~W=S\sqrt{\sum_{\substack{h=2}}^{2l}\left(\sum_{\substack{i}}^{M}K_{i}^{(h)}\sqrt{2}^{h}\Phi_{i}^{h}\right)^{2}}
\end{equation}
Here $\Phi_{i}$ are  the inflaton chiral superfields, $K_{i}^{(n)}$ are dimensionless coupling constants of their interactions and $K_{i}^{(h)}$ is the modified coupling constants in terms of the index $h$. The role of chiral multiplet $S$  and their allied issues have been explained in \cite{pinhero2017non-cano-con-attra-sing-inf}. One can notice that
this K\"ahler potential is invariant under the following transformation;
\begin{equation}\label{shift_symmetry}
\sum_{\substack{n}}^{l}K_{i}^{(n)}\Phi_{i}^{n}\rightarrow \sum_{\substack{n}}^{l}K_{i}^{(n)}\Phi_{i}^{n}+C_{i}^{l}
\end{equation}
This shift symmetry rather generalizes the shift symmetries proposed in \cite{kawasaki2000naturalchaotic,linear_running_kinetic_inflation_takahashi,running_kinetic_inflation_takahashi}, in the context of supergravity realization of the chaotic inflation and running kinetic inflation. Due to this shift symmetry Eq.(\ref{shift_symmetry}) the real component of the composite fields $\sum_{\substack{n}}^{l}K_{i}^{(n)}\Phi_{i}^{n}=\hat{\Phi_{i}}$, will be absent in the K\"ahler potential Eq.(\ref{eq_Kahler_potential}), and this real component $\text{Re}j\left[\sum_{\substack{n}}^{l}K_{i}^{(n)}\Phi_{i}^{n}\right]=\text{Re}\left[\hat{\Phi_{i}}\right]$  can be identified as the inflaton scalar fields. However physics is invariant under field transformation, one can also continue the same analysis  in terms of the real non-canonical variables. In such a scenario, real part of $\Phi_{i}$'s can be identified as inflatons  in terms of the non-canonical chiral fields $\Phi_{i}$ \cite{linear_running_kinetic_inflation_takahashi,running_kinetic_inflation_takahashi}. By decomposing $\Phi_{i}$ in terms of real and imaginary components $\Phi_{i}=\frac{1}{\sqrt{2}}\left(\phi_{i}+i\chi_{i}\right)$, and by assuming along flat direction $S=\chi_{i}=0$, The K\"ahler potential and superpotential defined in Eq.(\ref{eq_Kahler_potential})
can give the following Lagrangian in terms of the real fields
\begin{equation}
	L=\sqrt{-g}\left[\frac{R}{2}-\sum_{\substack{i}}^{M}\sum_{\substack{h=2}}^{2l}K_{i}^{(h)}\phi_{i}^{h-2}\frac{1}{2}\partial_{\mu}\phi_{i}\partial^{\mu}\phi_{i}-\sum_{\substack{h=2}}^{2l}\left(\sum_{\substack{i}}^{M}K_{i}^{(h)}\Phi_{i}^{h}\right)^{2}\right]
\end{equation}
Now we will add real conformon field $\chi$ into this theory with an equal footing in kinetic and potential term with respect to the $\phi$ field and consider a non-minimal conformal coupling to gravity in both $\phi$ and $\chi$ fields. It is quite well known that, ${\cal N}=1$ Poinca\'{r}e supergravity theory is not a conformally invariant theory, because of the lack of the conformon fields in it, which helps to make Einstein gravity part is conformally invariant. Hence, here the construction of our Lagrangian, from ${\cal N}=1$ supergravity theory  is purely phenomenological, since it includes the conformon fields also. So the total Lagrangian now reads in Jordan frame as
\begin{multline}\label{eq_main_Lagrangian_from_sugra}
	L=\sqrt{-g}\sum_{h=2}^{2l}  
	\left[\left(\frac{C^{(h)}\chi^{h}}{3h^{2}}
	-\sum_{i}^{M}\frac{K_{i}^{(h)}\phi_{i}^{h}}{3h^{2}}\right)R+ \frac{C^{(h)}\chi^{h-2}}{2}\partial_{\mu}
	\chi\partial^{\mu}\chi   
	- \sum_{i}^{M}\frac{K_{i}^{(h)}\phi_{i}^{h-2}}{2}\partial_{\mu}\phi_{i}
	\partial^{\mu}\phi_{i}  \right. \\  \left.
	-\frac{4}{9h^{4}}F\left ( \sum_{i}^{M}\frac{\phi_{i}}{\chi} \right )\left(\sum_{i}^{M}K_{i}^{(h)}\phi_{i}^{h}-C^{(h)}\chi^{h}\right)^{2} \right]
\end{multline}
where $C^{(h)}$ is the dimensionless coupling constants for the interaction of conformal field $\chi$. One can see that for $2l=2$ this Lagrangian Eq.(\ref{eq_main_Lagrangian_from_sugra}) boils down to the conformal invariant Lagrangian with canonical kinetic terms, which is used to study the construction of conformal multi-field inflation. By adding higher order terms into this theory, we here explicitly breaking this conformal symmetry. But one can notice that this Lagrangian has a conformal invariance if one avoids the summation in $h$ index, under the following set of transformations 
\begin{equation}\label{eq_conf_transformations_in_h_index}
	g^{'}_{\mu\nu}=e^{-2\sigma(x)}g_{\mu\nu},~~~~~~~~
	\chi^{'}=e^{\frac{2}{h}\sigma(x)}\chi,~~~~~~~~
	\phi^{'}=e^{\frac{2}{h}\sigma(x)}\phi
\end{equation}
Although this observation is irrelevant in the present discussion, one can pass through this kind of Lagrangian Eq.(\ref{eq_main_Lagrangian_from_sugra}) from a superconformal approach when the second conformal field say, $\eta$ is decouples from the theory, which we will show in the following discussion. In order to investigate such a possibility, one can start with defining a special kind of K\"ahler embedding manifold of the form
\begin{equation}\label{eq_N(X,X)}
{\cal N}(X, \bar{X})=\left| S \right|^{2}-3\varsigma \frac{(S\bar{S})^{2}}{\left |X_{1}^{0}  \right |^{2}-
	\left | X^{i} \right |^{2}}+\sum_{h=a+b}^{N}
\left(\sum_{i}^{M}\frac{K_{i}^{(h)}(X^{i})^{a}
	(\bar{X}^{\bar{i}})^{b}}{E^{(h)}(X_{2}^{0})^{a-1}
	(\bar{X_{2}}^{\bar{0}})^{b-1}}-
\frac{C^{(h)}(X_{1}^{0})^{a}(\bar{X_{1}}^{\bar{0}})^{b}}
{E^{(h)}(X_{2}^{0})^{a-1}(\bar{X_{2}}^{\bar{0}})^{b-1}}\right )
\end{equation}
where $X_{1}^{0}$ and $X_{2}^{0}$ are the complex scalar conformons, $X^{i}=\Phi_{i}$ are the inflaton superfields and $S$ is the sGoldstino field which serve as a stabilizer field. $K_{i}^{(h)}$, $C^{(h)}$, $E^{(h)}$ are the corresponding dimensionless coupling constants for inflaton fields, first and second conformal fields respectively, and the values of $K_{i}^{(2)}$, $C^{(2)}$, $E^{(2)}$ are normalized to 1. Our basic motivation is to introduce a second conformal scalar field $X_{2}^{0}$ in ${\cal N}(X,\bar{X})$, phenomenologically, to maintain the conformal symmetry in the Lagrangian to essentially deal with a  non-canonical non-homogeneous kinetic term for the both ghost and inflaton fields (i.e., with a powered summation in the Lagrangian ) and to maintain the homogeneity of the K$\ddot{a}$hler embedding manifold ${\cal N}(X,\bar{X})$.
Thus for the K\"ahler embedding manifold Eq.(\ref{eq_N(X,X)}) and for  a potential of the form
\begin{equation}\label{eq_V(X,X)}
V=\sum_{h=a+b}^{N}\frac{1}{9}F\left(\frac{\sum_{i}^{M}X^{i}}{X_{1}^{0}}\right)
\left(\sum_{i}^{M}\frac{K_{i}^{(h)}(X^{i})^{a}(\bar{X}^{\bar{i}})^{b}}
{E^{(h)}(X_{2}^{0})^{a-1}
	(\bar{X_{2}}^{\bar{0}})^{b-1}}-
\frac{C^{(h)}(X_{1}^{0})^{a}(\bar{X_{1}}^{\bar{0}})^{b}}
{E^{(h)}(X_{2}^{0})^{a-1}(\bar{X_{2}}^{\bar{0}})^{b-1}}\right )^{2}
\end{equation} 
the superconformal action for scalar-gravity part
\cite{kallosh_superconformal1}
\begin{equation}\label{eq_gen_sup_cnf_action}
\frac{1}{\sqrt{-g}}{\cal L}_{sc}^{\rm scalar-grav}=
-\frac{1}{6}{\cal N}(X,\bar{X})R-
G_{I\bar{J}}{\cal D}^{\mu}X^{I}{\cal D}_{\mu}\bar{X}^{\bar{J}}-
G^{I\bar{J}}{\cal W}_{I}\bar{\cal W}_{\bar{J}}
\end{equation}
will be conformal invariant under the following set of transformations
\begin{equation}\label{eq_conf_transformations}
g^{'}_{\mu\nu}=e^{-2\sigma(x)}g_{\mu\nu},~~~~~~~~
(X^{I})^{'}=e^{\sigma(x)}X^{I},~~~~~~~~
(\bar{X}^{\bar{J}})^{'}=e^{\sigma(x)}\bar{X}^{\bar{J}}
\end{equation}
With the advantage of the conformal symmetry Eq.(\ref{eq_conf_transformations}) one can gauge away the conformal fields which are negative in kinetic terms from the theory by fixing a gauge, since there are no degrees of freedom associated with these fields. The most convenient way of doing this gauge fixing is to choose a gauge which is ${\cal N}(X,\bar{X})=3$ in Eq.(\ref{eq_gen_sup_cnf_action}). This helps us to recover the standard Einstein term $-\frac{R}{2}$ of supergravity in Eq.(\ref{eq_gen_sup_cnf_action}). This gauge fixing can be interpreted as a migration from ${\cal N}=1$ superconformal theory to  ${\cal N}=1$ standard Poinca\'{r}e supergravity theory via spontaneous breaking of super conformal symmetry. For simplicity, we choose a gauge
\begin{equation}\label{eq_funda_gauge}
{\cal N}(X,\bar{X})=-3(N-1)
\end{equation}
(see \cite{pinhero2017non-cano-con-attra-sing-inf}).This dilatational gauge can be achieved by choosing a two set of system of $N-1$ equations as follows
\begin{equation}\label{eq_gauge_eq_1}
E^{(h)}(X_{2}^{0})^{a-1}
(\bar{X_{2}}^{\bar{0}})^{b-1}=\frac{h^{2}}{4}
\end{equation}
with a condition $ab\simeq\frac{h^{2}}{4}$ and
\begin{equation}\label{eq_guage_eq_2}
C^{(h)}(X_{1}^{0})^{a}(\bar{X_{1}}^{\bar{0}})^{b}-K_{i}^{(h)}(X^{i})^{a}
(\bar{X}^{\bar{i}})^{b}=\frac{3}{2}h^{2}
\end{equation}
and by assuming fields are real during inflation (corresponding partners are stabilized at zero during inflation) i.e., 
\begin{equation}
X_{1}^{0}=\bar{X}_{1}^{\bar{0}}=\frac{\chi}{\sqrt{2}},~~~~~~
X_{2}^{0}=\bar{X}_{2}^{\bar{0}}=\frac{\eta}{\sqrt{2}},~~~~~~ X^{i}=\bar{X}^{\bar{i}}=\frac{\phi_{i}}{\sqrt{2}}
\end{equation}
Using Eq.(\ref{eq_gauge_eq_1}) we gauge away the second conformal field $\eta$ from the theory and the Eq.(\ref{eq_gen_sup_cnf_action}) becomes,
\begin{multline}\label{eq_main_Lagrangian}
L=\sqrt{-g}\sum_{h=2}^{N}  
\left[\left(\frac{C^{(h)}\chi^{h}}{3h^{2}}
-\sum_{i}^{M}\frac{K_{i}^{(h)}\phi_{i}^{h}}{3h^{2}}\right)R+ \frac{C^{(h)}\chi^{h-2}}{2}\partial_{\mu}
\chi\partial^{\mu}\chi   
- \sum_{i}^{M}\frac{K_{i}^{(h)}\phi_{i}^{h-2}}{2}\partial_{\mu}\phi_{i}
\partial^{\mu}\phi_{i}  \right. \\  \left.
-\frac{4}{9h^{4}}F\left ( \sum_{i}^{M}\frac{\phi_{i}}{\chi} \right )\left(\sum_{i}^{M}K_{i}^{(h)}\phi_{i}^{h}-C^{(h)}\chi^{h}\right)^{2} \right]
\end{multline}
Consequently, this Lagrangian coincides with the theory  Eq.(\ref{eq_main_Lagrangian_from_sugra}) for $N=2l$.
Thus we end up with a Lagrangian which contains non-canonical non-homogeneous kinetic terms for both conformal field $\chi$ and for inflaton fields $\phi_{i}$ with an equal footing, after the decoupling of the second conformal field $\eta$ from the superconformal action. Now the role of this second conformal field $X_{2}^{0}=\frac{\eta}{2}$ is clear, which helps us to maintain the conformal symmetry in the theory when one essentially wants to deal with a non-canonical non-homogeneous kinetic term for the conformal field $\chi$.


\section{Non-canonical multi-field inflationary attractors}\label{Non-canonical multi-field inflationary attractors}
In this section we start our analysis by using the Lagrangian Eq.(\ref{eq_main_Lagrangian}). 
Note that this Lagrangian Eq.(\ref{eq_main_Lagrangian}) has an enhanced conformal symmetry when the parameter $N$ takes the value 2 and will reduce to canonical multi-field model, which is the prime Lagrangian that studied in the construction of multi-field conformal attractors discussed in the literature \cite{kallosh2013multi}. 
In order to explore the multi-field scenario in the Lagrangian Eq.(\ref{eq_main_Lagrangian}) in a more simpler way,
we consider only two orthogonal fields:
\begin{equation}
\phi_{1}=\rho\cos\theta~~~~~~ \text{and}~~~~~~~~\phi_{2}=\rho\sin\theta.
\end{equation}
In terms of this newly defined radial and angular fields, $\rho$ and $\theta$ respectively, the above Lagrangian Eq.(\ref{eq_main_Lagrangian}) looks
\begin{multline}
L^{J}=\sqrt{-g}\sum_{h=2}^{N}\left[\frac{C^{(h)}\chi^{h-2}}{2}\partial_{\mu}
\chi\partial^{\mu}\chi  -\frac{1}{8}A^{(h-4)}(\theta)\sin^{2}2\theta\rho^{h}\partial_{\mu}\theta\partial^{\mu}\theta + \frac{1}{2}A^{(h-2)}(\theta)\sin2\theta\rho^{h-1}\right. \\  \left. \times\partial_{\mu}\rho\partial^{\mu}\theta- \frac{1}{2}A^{(h)}(\theta)\rho^{h-2}\partial_{\mu}\rho\partial^{\mu}\rho +\left(C^{(h)}\chi^{h}- A^{(h)}(\theta)\rho^{h}\right )\frac{R}{3h^{2}}\right. \\  \left. -\frac {4}{9h^{4}}F\left(\frac{\rho}{\chi},\theta\right)\left(A^{(h)}(\theta)\rho^{h}- C^{(h)}\chi^{h}\right )^{2} \right]
\end{multline}
where
\begin{equation}
A^{(h)}(\theta)=K_{1}^{(h)}\cos^{h}\theta+K_{2}^{(h)}\sin^{h}\theta 
\end{equation}
\begin{equation}
A^{(h-2)}(\theta)=K_{1}^{(h)}\cos^{h-2}\theta-K_{2}^{(h)}\sin^{h-2}\theta 
\end{equation}
\begin{equation}
A^{(h-4)}(\theta)=K_{1}^{(h)}\cos^{h-4}\theta+K_{2}^{(h)}\sin^{h-4}\theta 
\end{equation}
Now we will gauge away the first conformal compensator field $\chi$ from the theory by using the Eq.(\ref{eq_guage_eq_2}). With the use of these newly defined radial and angular fields the Eq.(\ref{eq_guage_eq_2}) becomes
\begin{equation}\label{eq_gauge_eq_2_for_2_fields}
C^{(h)}\chi^{h}- A^{(h)}(\theta)\rho^{h}=\frac{3h^{2}}{2}
\end{equation}
Each of these equations represent the hyperboloid of two sheets provided with a constraint $0<\theta<\frac{\pi}{2}$, and these can be represented in terms of canonically normalized fields $\psi_{h-1}$ as
\begin{equation}\label{eq_for_chi}
\chi=\left [\frac{3h^{2}}{2C^{(h)}} 
\right ]^{\frac{1}{h}}\cosh ^{\frac{2}{h}}
\left ( \frac{\psi_{h-1}}{\sqrt{6}} \right )
\end{equation}
\begin{equation}\label{eq_for_rho}
\rho=\left [ \frac{3h^{2}}{2A^{(h)}(\theta)} 
\right ]^{\frac{1}{h}}\sinh ^{\frac{2}{h}}
\left ( \frac{\psi_{h-1}}{\sqrt{6}} \right )
\end{equation}
Using the constraint Eq.(\ref{eq_gauge_eq_2_for_2_fields}) and in terms of the newly redefined canonically normalized fields $\psi_{h-1}$ the original Lagrangian Eq.(\ref{eq_main_Lagrangian}) can be expressed in Einstein frame
\begin{multline}\label{eq_lagrangian_psi_(h-1)}
L^{E}=\sqrt{-g}\sum_{h=2}^{N}\left \{ \frac{R}{2}-\frac{1}{2}\partial_\mu\psi_{h-1}\partial^{\mu}\psi_{h-1} +\frac{3h^{2}}{16}\left[\left( \frac{A^{(h-2)}(\theta)}{A^{(h)}(\theta)}\right )^{2}-\frac{A^{(h-4)}(\theta)}{A^{(h)}(\theta)} \right ]\sin^{2}\theta\right. \\  \left.\times\sinh^{2}\left(\frac{\psi_{h-1}}{\sqrt{6}}\right )\partial_{\mu}\theta\partial^{\mu}\theta-F\left[\left(\frac{C^{(h)}}{A^{(h)}(\theta)}\right)^{\frac{1}{h}}\tanh^{\frac{2}{h}}\left(\frac{\psi_{h-1}}{\sqrt{6}} \right ),\theta \right ]  \right \}
\end{multline}
From the equations Eq.(\ref{eq_for_chi}) and Eq.(\ref{eq_for_rho}), one can readily see that these $\psi_{h-1}$ fields are dependent on each other so one can easily express all these fields in terms of a single field $\psi$ as:
\begin{equation}
\psi_{h-1}=\sqrt{6}\sinh^{-1}\left[
\frac{(\sqrt{6})^{\frac{h}{2}}}{h}\sqrt{\frac{2A^{(h)}(\theta)}{3}}
\sinh^{\frac{h}{2}}\left(\frac{\psi}{\sqrt{6}} \right ) \right ]
\end{equation}
substitute this field back into the Lagrangian Eq.(\ref{eq_lagrangian_psi_(h-1)}), we will essentially end up with a non-canonical kinetic terms for both radial and angular fields. As a result final Lagrangian takes the form
\begin{multline}\label{eq_multifield_final_lagrangian}
L^{E}=\sqrt{-g}\sum_{h=2}^{N}\left \{ \frac{R}{2}-\frac{\left(\sqrt{6}\right)^{h}A^{(h)}(\theta)
	\sinh^{h-2}\left(\frac{\psi}{\sqrt{6}}\right)
	\cosh^{2}\left(\frac{\psi}{\sqrt{6}}\right)}{12\left[1+
	\frac{2\left(\sqrt{6}\right)^{h}}{3h^{2}}A^{(h)}(\theta)\sinh^{h}
	(\frac{\psi}{\sqrt{6}})\right]}\partial_{\mu}\psi
\partial^{\mu}\psi \right. \\  \left. +\frac{\left(\sqrt{6}\right)^{h-1}A^{(h-2)}(\theta)\sin2\theta\sinh^{h-1}\left(\frac{\psi}{\sqrt{6}}\right)
	\cosh\left(\frac{\psi}{\sqrt{6}}\right)}{2\left[1+
	\frac{2\left(\sqrt{6}\right)^{h}}{3h^{2}}A^{(h)}(\theta)\sinh^{h}
	\left(\frac{\psi}{\sqrt{6}}\right)\right]}\partial_{\mu}\psi
\partial^{\mu}\theta \right. \\  \left. +\left[\frac{\left[A^{(h-2)}(\theta) \right ]^{2}}{A^{(h)}(\theta)}-A^{(h-4)}(\theta) -\frac{\left[A^{(h-2)}(\theta)\right]^{2}
}{A^{(h)}(\theta)\left[1+
	\frac{2\left(\sqrt{6}\right)^{h}}{3h^{2}}A^{(h)}(\theta)\sinh^{h}
	\left(\frac{\psi}{\sqrt{6}}\right)\right]}\right ]\frac{\left(\sqrt{6}\right)^{h}}{8}\sin^{2}2\theta\right. \\  \left.\times\sinh^{h}\left(\frac{\psi}{\sqrt{6}} \right )\partial_{\mu}\theta
\partial^{\mu}\theta-
F\left (\tanh ^{\frac{2}{h}}\left [ \sinh^{-1}\left[\frac{\left(\sqrt{6}\right)^{\frac{h}{2}}}{h}
\sqrt{\frac{2A^{(h)}(\theta)}{3}}\sinh^{\frac{h}{2}}
\left(\frac{\psi}{\sqrt{6}} \right ) 
\right ] \right ],\theta\right )\right \}
\end{multline}
Here we have assumed $C^{(h)}=A^{(h)}(\theta)$ for the simplicity. Thus our theory involves two non-canonically normalized scalar fields $\psi$ and $\theta$ and their kinetic mixing term with the Einstein frame potential 
\begin{equation}\label{eq_final_functional_potential}
V\left(\psi, \theta\right)=\sum_{h=2}^{N}F\left (\tanh ^{\frac{2}{h}}\left [ \sinh^{-1}\left[\frac{\left(\sqrt{6}\right)^{\frac{h}{2}}}{h}
\sqrt{\frac{2A^{(h)}(\theta)}{3}}\sinh^{\frac{h}{2}}
\left(\frac{\psi}{\sqrt{6}} \right ) 
\right ] \right ],\theta\right )
\end{equation}
\begin{figure}
	\centering
	\begin{subfigure}{.5\textwidth}
		\centering
		\includegraphics[width=.9\linewidth]{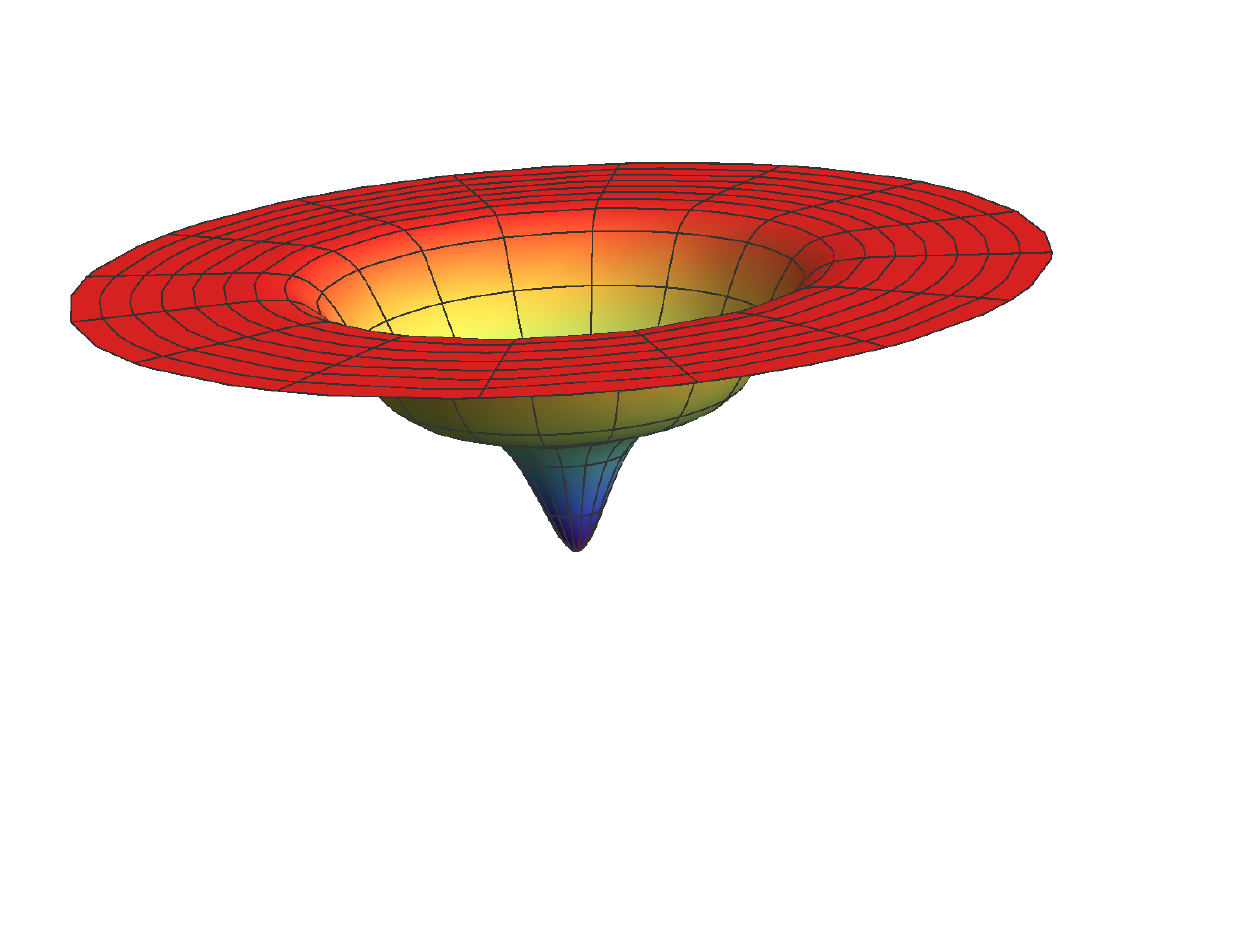}
		\caption{Modulated T-model potential for the Eq.(\ref{eq_final_functional_potential}) without $\theta$ dependence and with $N=3$}
		\label{fig:non-canonical T-model}
	\end{subfigure}%
	\begin{subfigure}{.5\textwidth}
		\centering
		\includegraphics[width=.9\linewidth]{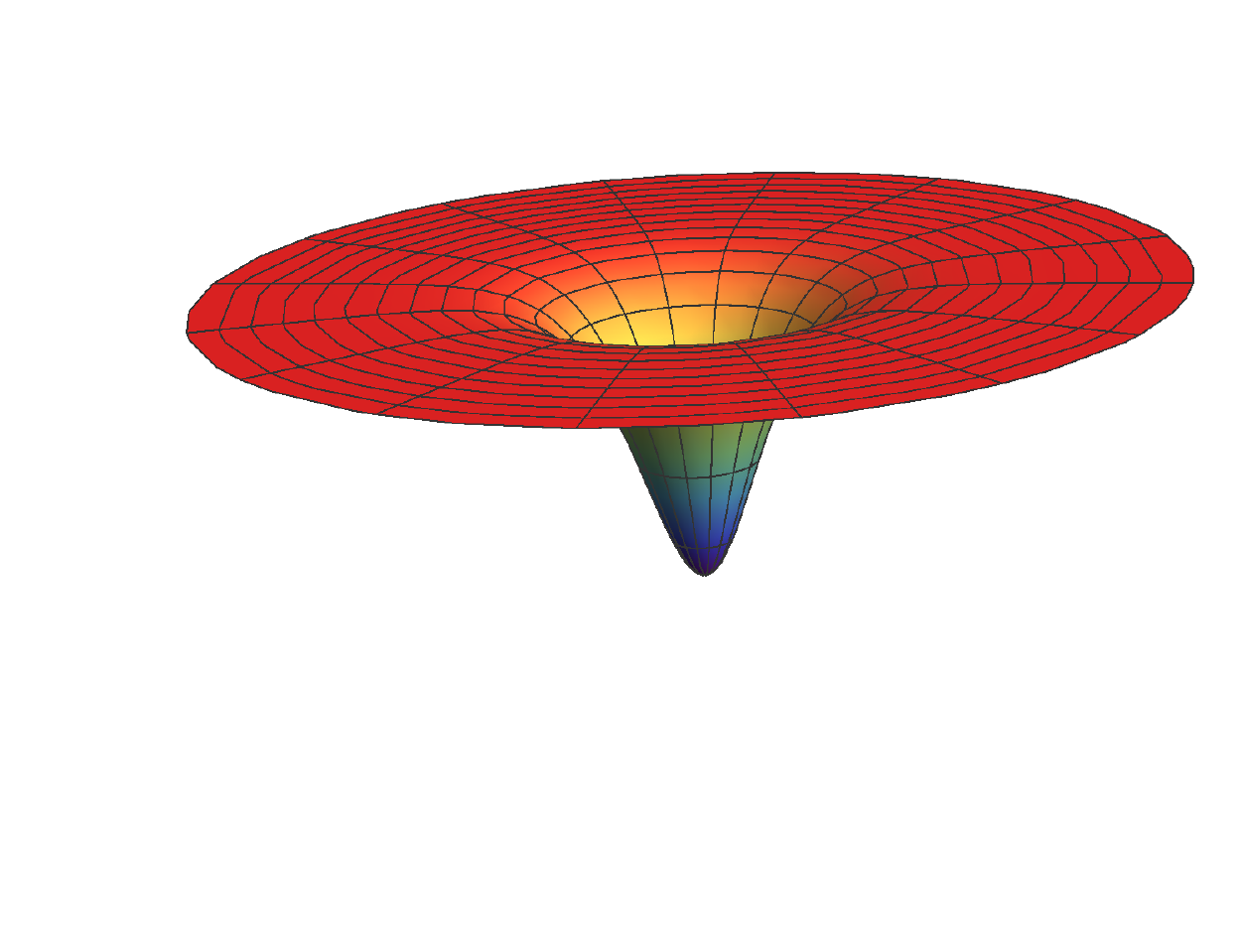}
		\caption{T-model potential, $V=\tanh^{2}\frac{\psi}{\sqrt{6}}$}
		\label{fig:canonical T-model}
	\end{subfigure}
	\caption{T-model potentials in both non-canonical Fig.(\ref{fig:non-canonical T-model}) and canonical Fig.(\ref{fig:canonical T-model}) conformal attractors scenario.}
	\label{fig:T-model}
\end{figure} 
Now one can notice that for $N=2$ case, this Lagrangian Eq.(\ref{eq_multifield_final_lagrangian}) reduces to a theory with a canonically normalized $\psi$ field and a non-canonically normalized $\theta$ field with the Einstein frame potential $V=F\left(\tanh\frac{\psi}{\sqrt{6}},\theta\right)$, which is known as the multi-field cosmological attractors and which has been well studied in the literature \cite{kallosh2013multi}. Also, if one further assumes the potential Eq.(\ref{eq_final_functional_potential}) is the function of radial field $\psi$ only, then the Eq.(\ref{eq_multifield_final_lagrangian}) will end up with essentially single field non-canonically normalized field $\psi$ and the theory is known as non-canonical single field conformal attractors \cite{pinhero2017non-cano-con-attra-sing-inf} and from here it is possible to recover a modulated T-model potential as shown in the Fig.(\ref{fig:non-canonical T-model}). Also if the potential Eq.(\ref{eq_final_functional_potential}) is only the function of $\psi$ field and the value of $N$ is restricted to 2 in Eq.(\ref{eq_multifield_final_lagrangian}), this will boils down to canonical conformal attractor scenario \cite{kallosh2013universality} and from here one can recover T-model potential as shown in the Fig.(\ref{fig:canonical T-model}). Thus our theory generalizes all these models in  a single framework. 

\begin{figure}
	\centering
	\begin{subfigure}{.5\textwidth}
		\centering
		\includegraphics[width=.8\linewidth]{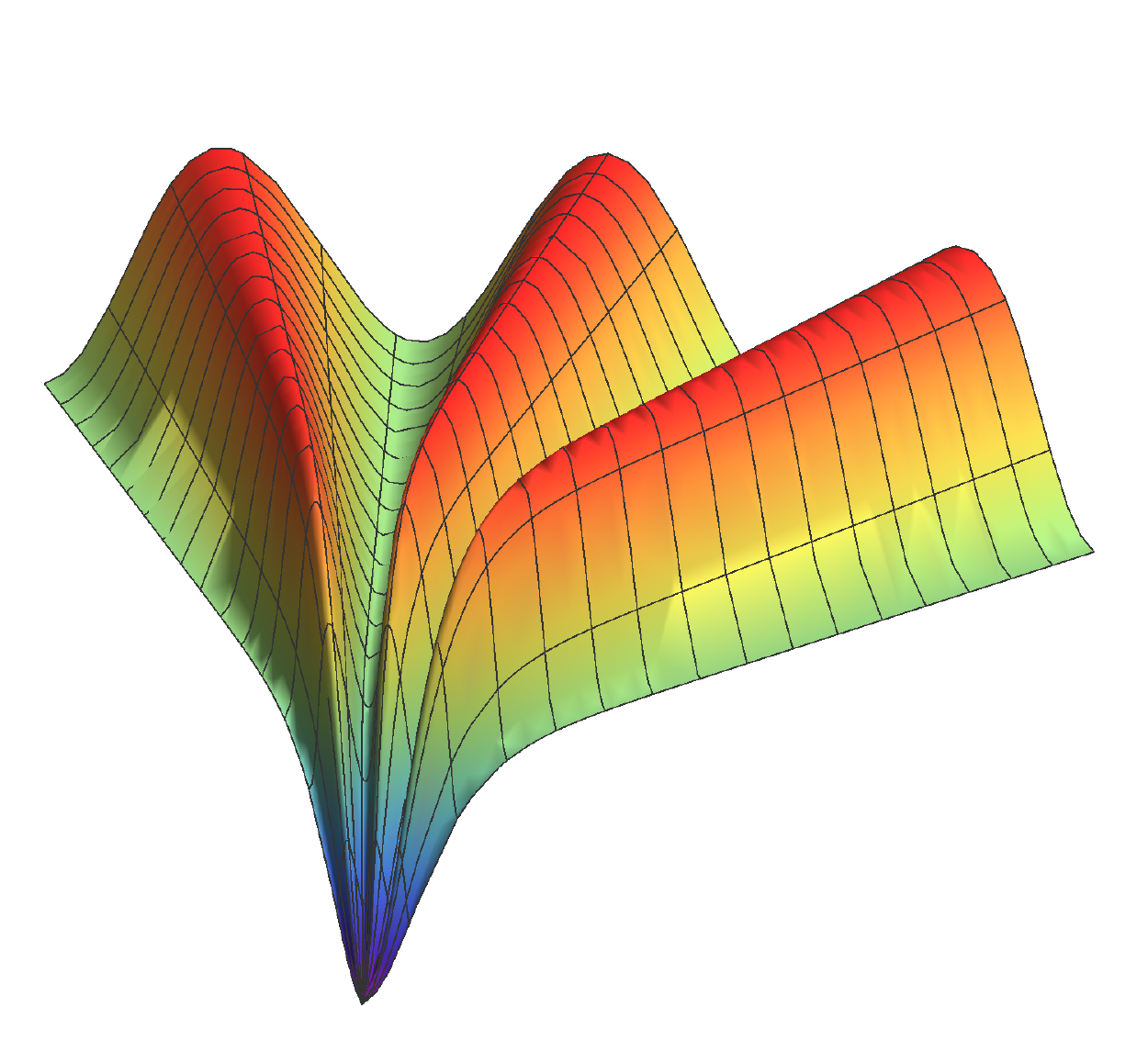}
		\caption{for $N=2$}
		\label{fig:sub1}
	\end{subfigure}%
	\begin{subfigure}{.5\textwidth}
		\centering
		\includegraphics[width=.8\linewidth]{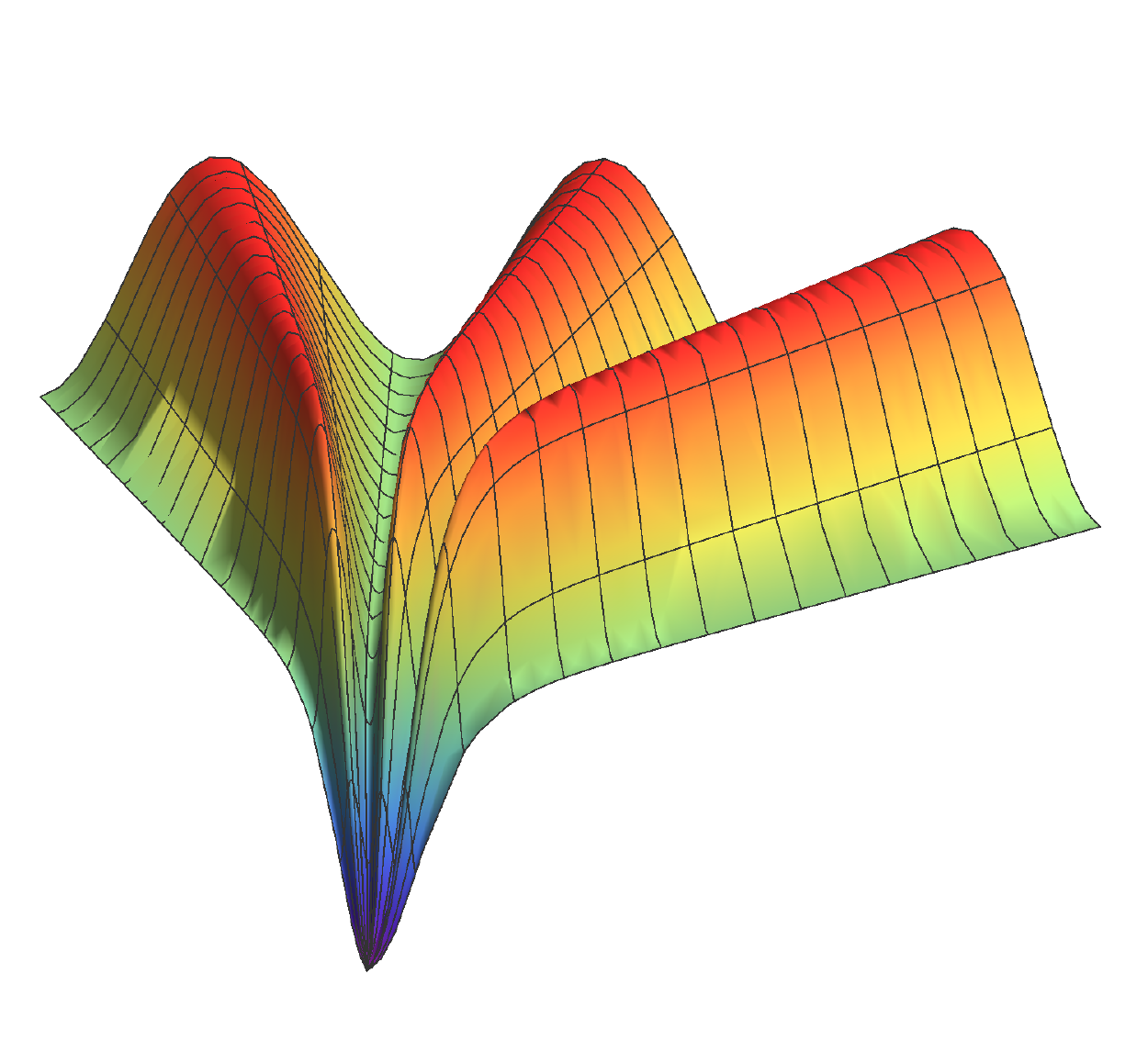}
		\caption{for $N=3$}
		\label{fig:sub2}
	\end{subfigure}
	\begin{subfigure}{.5\textwidth}
		\centering
		\includegraphics[width=.8\linewidth]{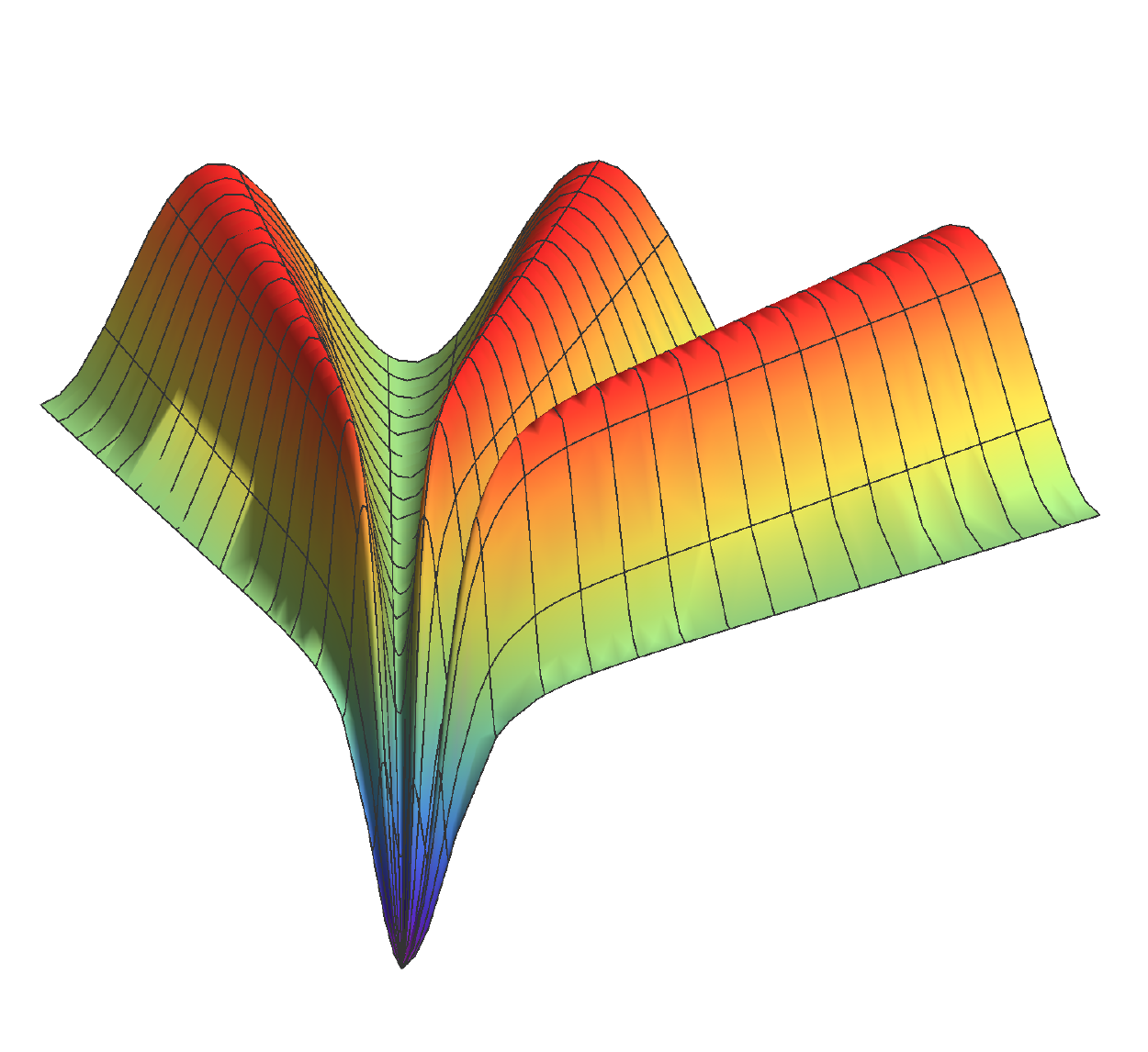}
		\caption{for $N=4$}
		\label{fig:sub3}
	\end{subfigure}%
	\begin{subfigure}{.5\textwidth}
		\centering
		\includegraphics[width=.8\linewidth]{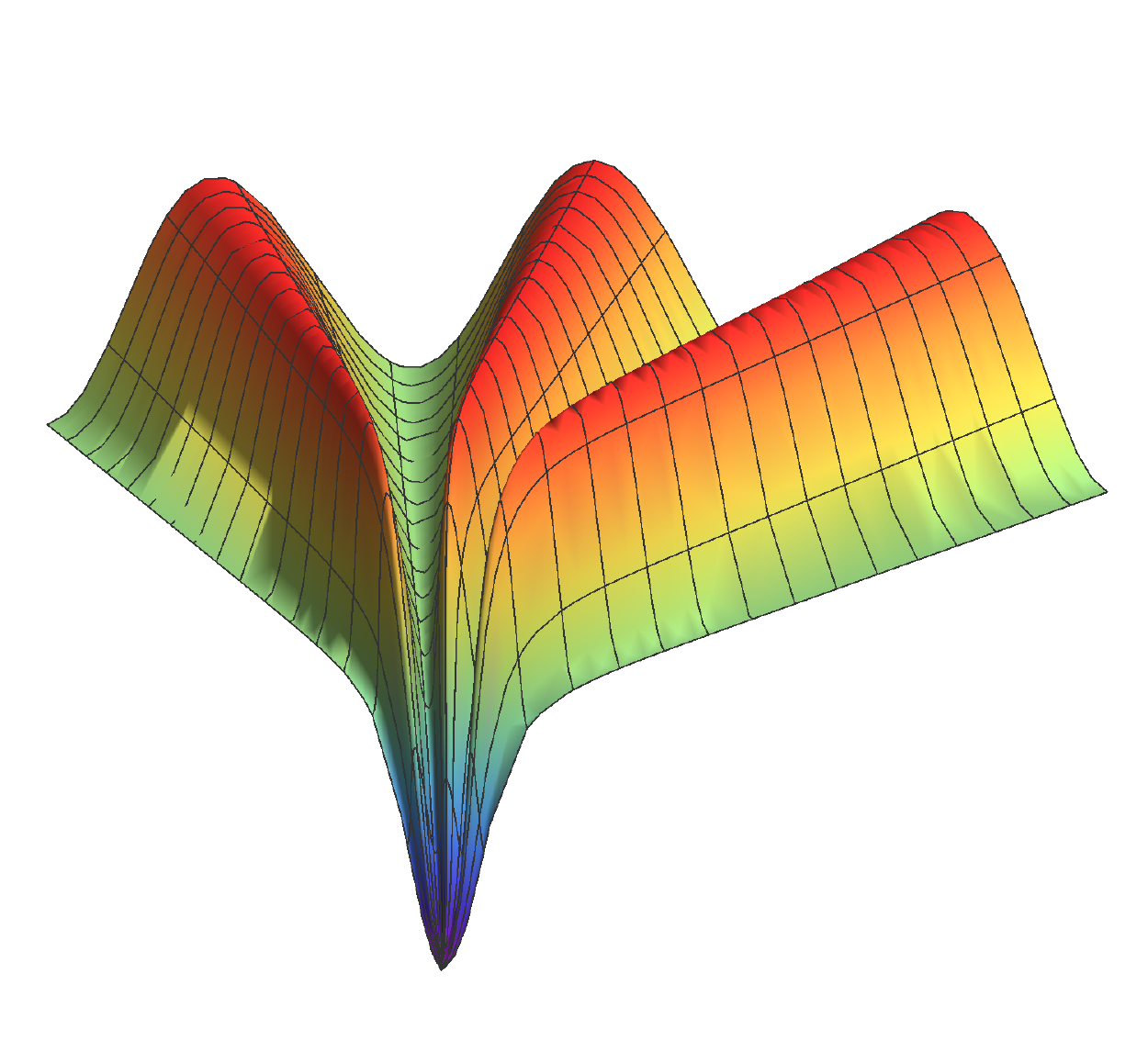}
		\caption{for $N=5$}
		\label{fig:sub4}
	\end{subfigure}
	\caption{Figure depicts the bird's eye view of the potential defined in Eq.(\ref{eq_toy_potential_1}) for the values $n^{'}=1$, $g=3$ in the quadrant $0<\theta<\frac{\pi}{2}$, and for all coupling constants we choose $K_{1}^{(h)}=K_{2}^{(h)}=1$ for the comparison. }
	\label{fig_toy_potential_1}
\end{figure}

However, in general, the potential in the Lagrangian Eq.(\ref{eq_multifield_final_lagrangian}) should depend on the angular field $\theta$ also, and as a result one can expect ridges and valleys along the radial direction of Fig.(\ref{fig:T-model}). A simple example is shown in the Fig.(\ref{fig_toy_potential_1}) for the potential of the form
\begin{equation}\label{eq_toy_potential_1}
V=\sum_{h=2}^{N}\left(A+B\sin^{2}2g\theta\right)\tanh ^{\frac{4n^{'}}{h}}\left \{ \sinh^{-1}\left[\frac{\left(\sqrt{6}\right)^{\frac{h}{2}}}{h}
\sqrt{\frac{2A^{(h)}(\theta)}{3}}\sinh^{\frac{h}{2}}
\left(\frac{\psi}{\sqrt{6}} \right ) 
\right ] \right\}
\end{equation}
and, as a result, the theory then represents an intrinsically multi-field non-canonical conformal attractor scenario. Below we will study these possibilities, i.e., we will discuss the trajectory and inflation dynamics for these fields along the ridges and valleys of the potential with the use of non-canonical kinetic terms of the action defined in Eq.(\ref{eq_multifield_final_lagrangian}).

\section{Inflationary dynamics}\label{inflationary_dynamics}
\begin{figure}
	\centering
	\begin{subfigure}{.5\textwidth}
		\centering
		\includegraphics[width=.8\linewidth]{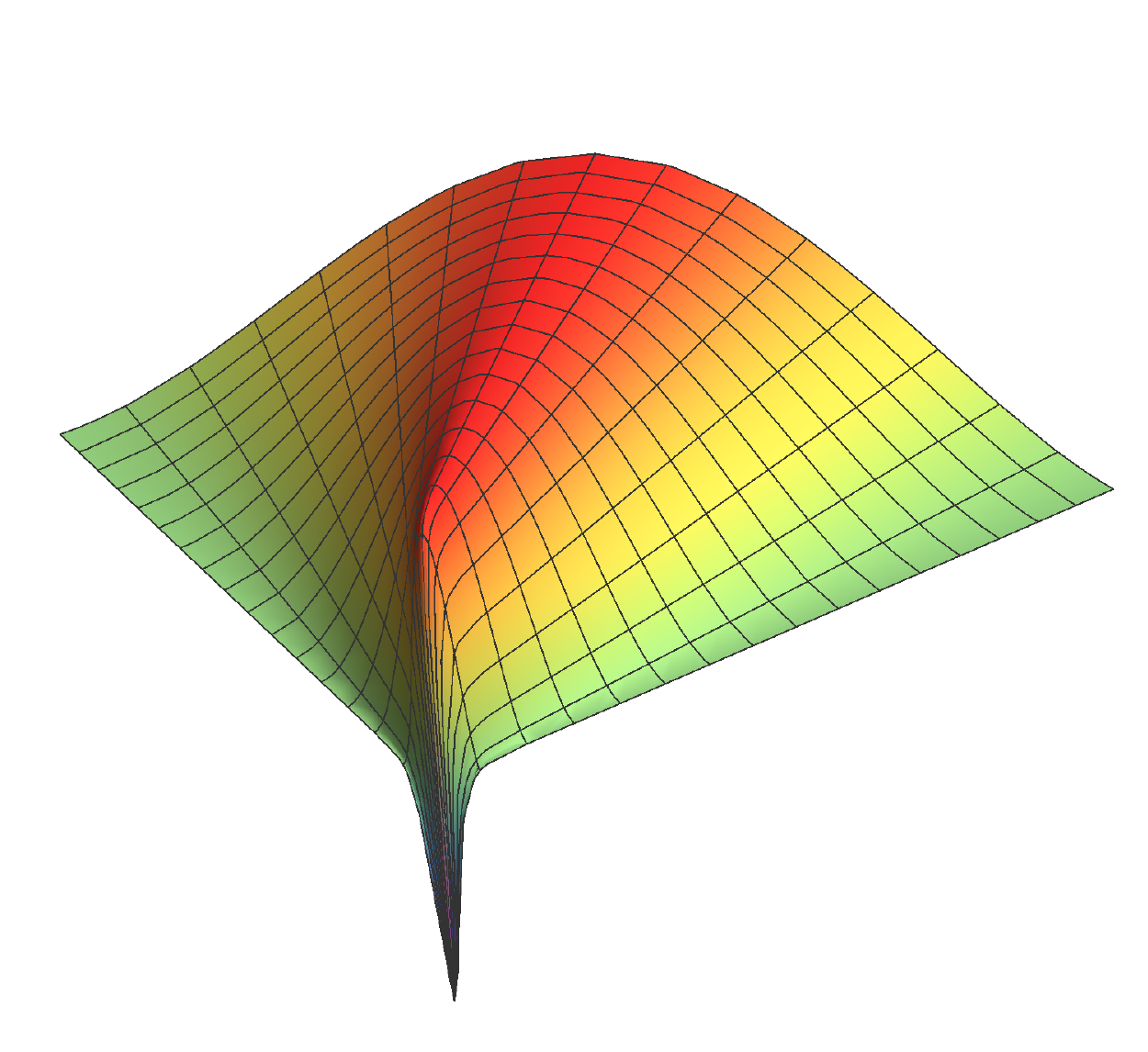}
		\caption{for $N=2$}
		\label{fig1}
	\end{subfigure}%
	\begin{subfigure}{.5\textwidth}
		\centering
		\includegraphics[width=.8\linewidth]{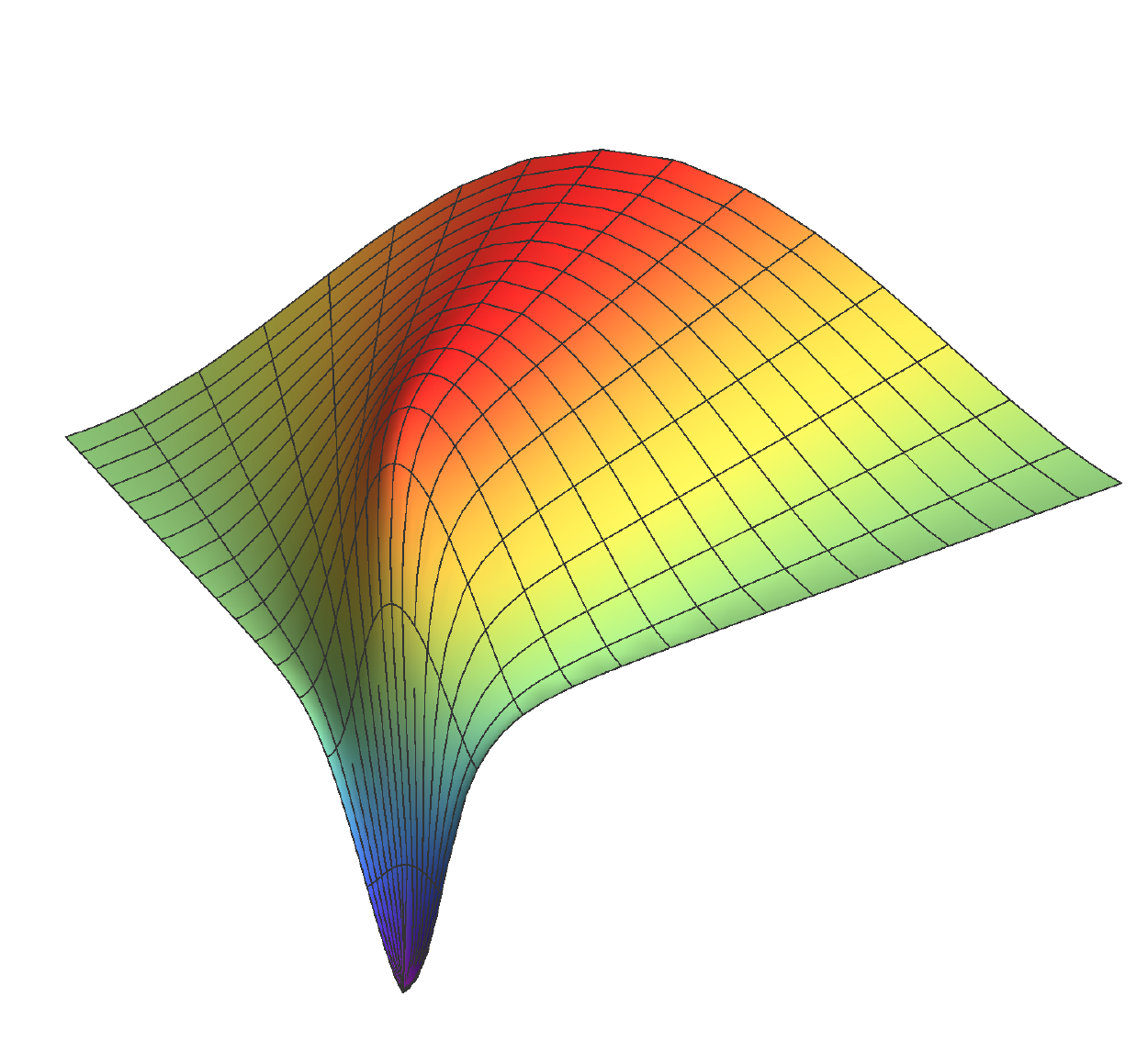}
		\caption{for $N=3$}
		\label{fig2}
	\end{subfigure}
	\begin{subfigure}{.5\textwidth}
		\centering
		\includegraphics[width=.8\linewidth]{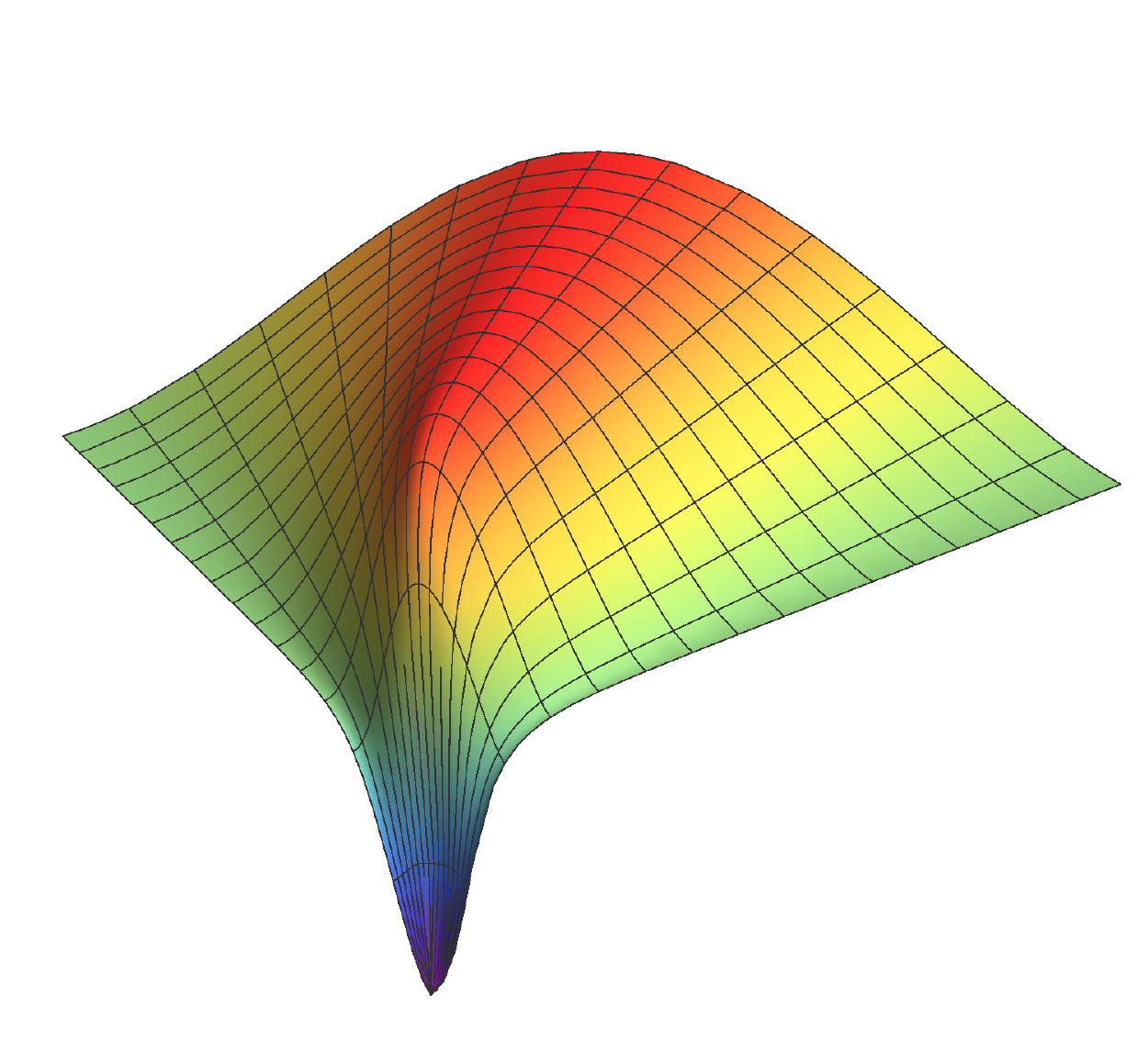}
		\caption{for $N=4$}
		\label{fig3}
	\end{subfigure}%
	\begin{subfigure}{.5\textwidth}
		\centering
		\includegraphics[width=.8\linewidth]{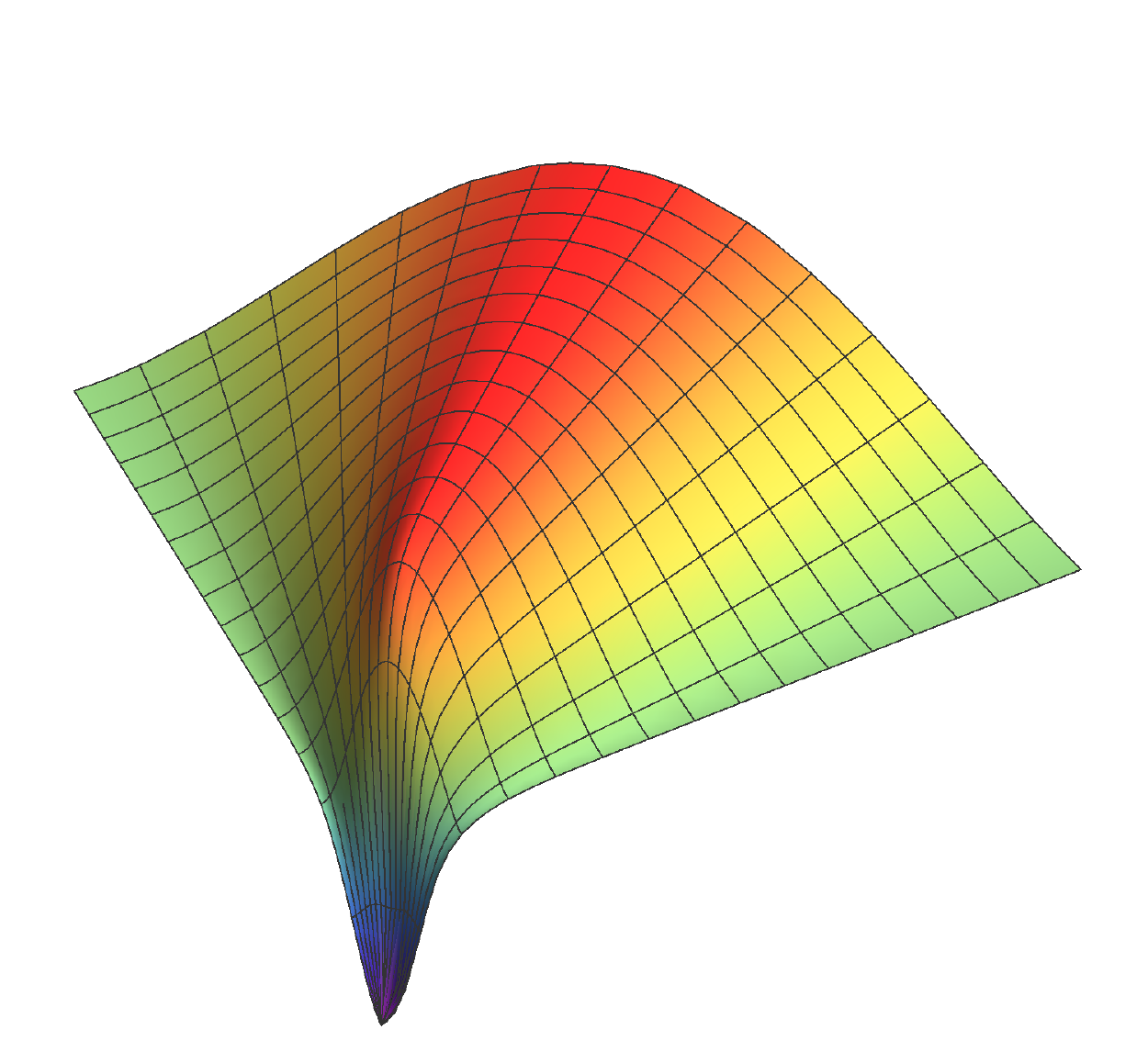}
		\caption{for $N=5$}
		\label{fig4}
	\end{subfigure}
	\caption{Figure depicts the bird's eye view of the potential defined in Eq.(\ref{toy_potential_1_exact_form}) for the values $A=B$ in the quadrant $0<\theta<\frac{\pi}{2}$,  and for all coupling constants we choose $K_{1}^{(h)}=K_{2}^{(h)}=1$ for the comparison. }
	\label{fig_toy_potential_2}
\end{figure}

By varying the action Eq.(\ref{eq_multifield_final_lagrangian}), with respect to $\psi$ and with respect to $\theta$ one can calculate the equation of motion for the fields $\psi$ and $\theta$ respectively [see Eq.(\ref{equ_of_motion_for_psi_field}) and Eq.(\ref{equ_of_motion_for_theta_field})]. As we are interested to study the evolution of these fields at large $\psi$ values i.e., at $\psi>>1$  and for slow-roll regime these equation of motion for the fields can be approximately written as

\begin{multline}\label{very_approximated_eq_motion for_psi_field}
\left\{1+\left[ \frac{N(N+1)(2N+1)-30}{24}\right ]\left(1+4e^{-\sqrt{\frac{2}{3}}\psi} \right )\right\} 3H\dot{\psi}\\-\frac{3\sin2\theta}{4\sqrt{6}}\left(1+2e^{-\sqrt{\frac{2}{3}}\psi} \right )3H\dot{\theta}\sum_{h=2}^{N}\frac{h^{2}A^{(h-2)}(\theta)}{A^{(h)}(\theta)}\\+\sum_{h=2}^{N}\left(\frac{\left[A^{(h-2)}(\theta) \right ]^{2}}{A^{(h)}(\theta)}-A^{(h-4)}(\theta) \right )\frac{(\sqrt{6})^{h-1}}{8}\frac{h\sin^{2}2\theta }{2^{h}}e^{\frac{h\psi}{\sqrt{6}}}\\\times\left[1+(2-h)e^{-\sqrt{\frac{2}{3}}\psi} \right ]\dot{\theta}^{2}\simeq-V_{\psi}
\end{multline}
and 
\begin{equation}\label{very_approximated_eq_motion for_theta_field}
3H\dot{\theta}\simeq\frac{-V_{\theta}}{\sum_{h=2}^{N}\left(A^{(h-4)}(\theta)-\frac{\left[A^{(h-2)}(\theta) \right ]^{2}}{A^{(h)}(\theta)} \right )\frac{(\sqrt{6})^{h}}{4}\frac{\sin^{2}2\theta }{2^{h}}e^{\frac{h\psi}{\sqrt{6}}}\left(1-he^{-\sqrt{\frac{2}{3}}\psi} \right )}
\end{equation}
In this approximation we express all hyperbolic functions in terms of exponential functions and keep only up to the terms $e^{-\sqrt{\frac{2}{3}}\psi}$ and neglect the higher order terms. 
For the slow-roll regime one can also calculate Friedmann's Equation for this theory as
\begin{equation}\label{approximated_friedmann_equation}
3H^{2}\simeq V\left(\psi, \theta\right)
\end{equation}
Exact form of this Friedmann's Equation, calculation of the slow-roll regime and the calculations of Eq.(\ref{very_approximated_eq_motion for_psi_field}) and Eq.(\ref{very_approximated_eq_motion for_theta_field}) have been shown in the appendix \ref{appendix_1}. Using Eq.(\ref{approximated_friedmann_equation}) and Eq.({\ref{very_approximated_eq_motion for_theta_field}}) one can write velocity of the angular field $\theta$  as
\begin{equation}\label{speed_for_theta_field}
\frac{\dot{\theta}}{H}\simeq-\frac{V_{\theta}}{V\sum_{h=2}^{N}\left(A^{(h-4)}(\theta)-\frac{\left[A^{(h-2)}(\theta) \right ]^{2}}{A^{(h)}(\theta)} \right )\frac{(\sqrt{6})^{h}}{4}\frac{\sin^{2}2\theta }{2^{h}}e^{\frac{h\psi}{\sqrt{6}}}\left(1-he^{-\sqrt{\frac{2}{3}}\psi} \right ) }
\end{equation}
From this equation one can observe that, as the value of $N$ increases, speed of $\theta$ field will be hugely suppressed. Thus the value of $N$ has a crucial role in determining the speed of these fields. As a result, one may question whether this value of $N$ is arising from the potential or from the non-canonical terms or a combination of both from the Eq.(\ref{eq_multifield_final_lagrangian}). In order to answer to this question and to understand the combined evolution of the fields during inflation, we further consider the potential of the form
\begin{equation}\label{toy_potential_1}
V=\sum_{h=2}^{N}F\left(z_{i}\right)=\sum_{h=2}^{N}A\left(z_{1}^{2}+z_{2}^{2}\right)+4Bz_{1}^{2}z_{2}^{2}
\end{equation}
where $z_{i}=\frac{\sum_{i}\phi_{i}}{\chi}$.
With the use of Eq.(\ref{eq_for_chi}) and Eq.(\ref{eq_for_rho}) the above potential takes the form
\begin{multline}\label{toy_potential_1_exact_form}
V=\sum_{h=2}^{N}\left\{A\tanh ^{\frac{4}{h}}\left [ \sinh^{-1}\left[\frac{\left(\sqrt{6}\right)^{\frac{h}{2}}}{h}
\sqrt{\frac{2A^{(h)}(\theta)}{3}}\sinh^{\frac{h}{2}}
\left(\frac{\psi}{\sqrt{6}} \right ) 
\right ] \right]\right.\\ \left.+B\sin^{2}2\theta\tanh ^{\frac{8}{h}}\left [ \sinh^{-1}\left[\frac{\left(\sqrt{6}\right)^{\frac{h}{2}}}{h}
\sqrt{\frac{2A^{(h)}(\theta)}{3}}\sinh^{\frac{h}{2}}
\left(\frac{\psi}{\sqrt{6}} \right ) 
\right ] \right]\right\}
\end{multline}
which is shown in the Fig.(\ref{fig_toy_potential_2}). For large values of $\psi\gg1$ this potential can be approximated as
\begin{equation}\label{toy_potential_approximated_for_large_psi}
V=A\left[\left(N-1\right)-4e^{-\sqrt{\frac{2}{3}}\psi}\right]+B\sin^{2}2\theta\left[\left(N-1\right)-8e^{-\sqrt{\frac{2}{3}}\psi}\right]
\end{equation}
\begin{figure}
	\centering
	\begin{subfigure}{.5\textwidth}
		\centering
		\includegraphics[width=.8\linewidth]{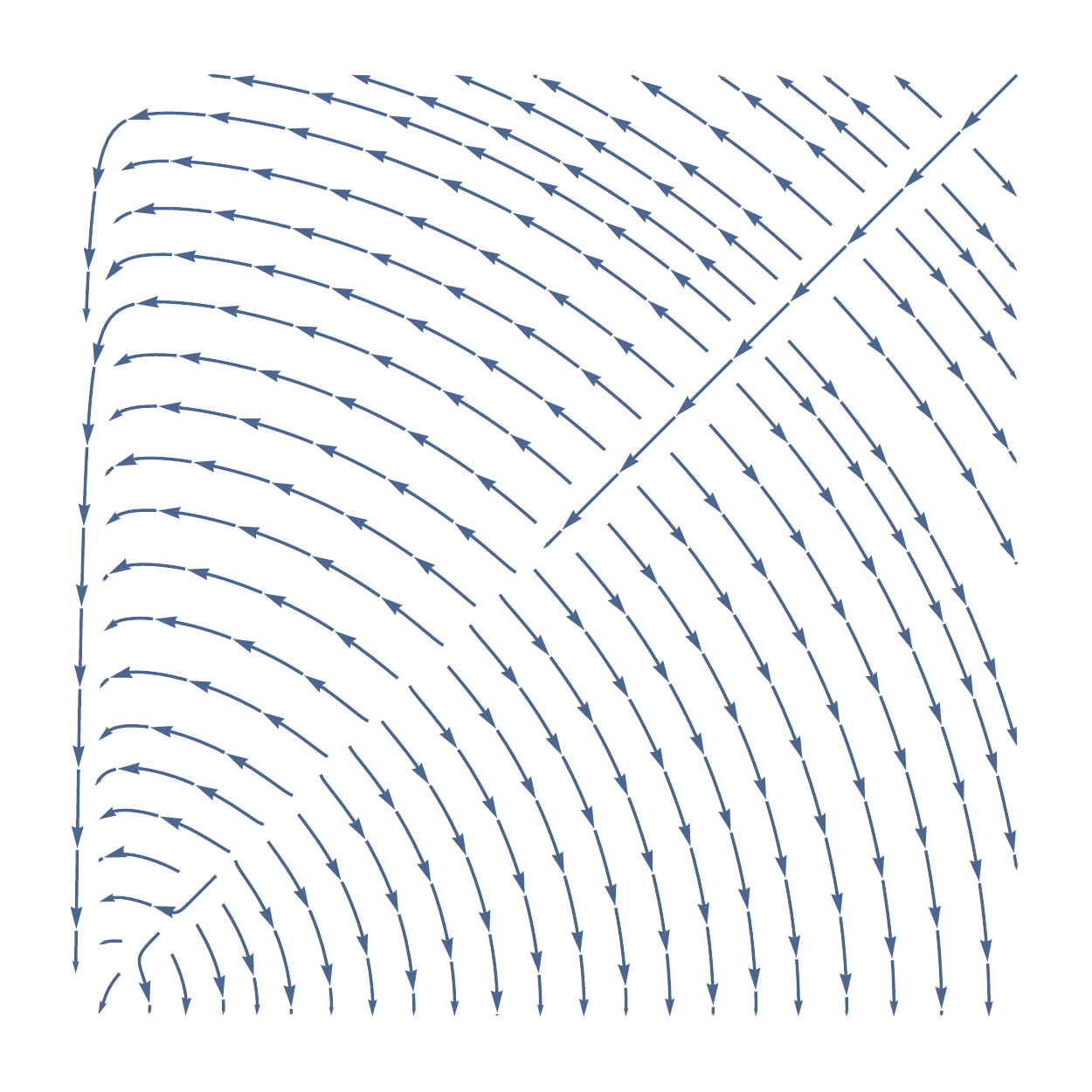}
		\caption{For $N=2$}
		\label{fig:for_N=2_psi=40_earlier}
	\end{subfigure}%
	\begin{subfigure}{.5\textwidth}
		\centering
		\includegraphics[width=.8\linewidth]{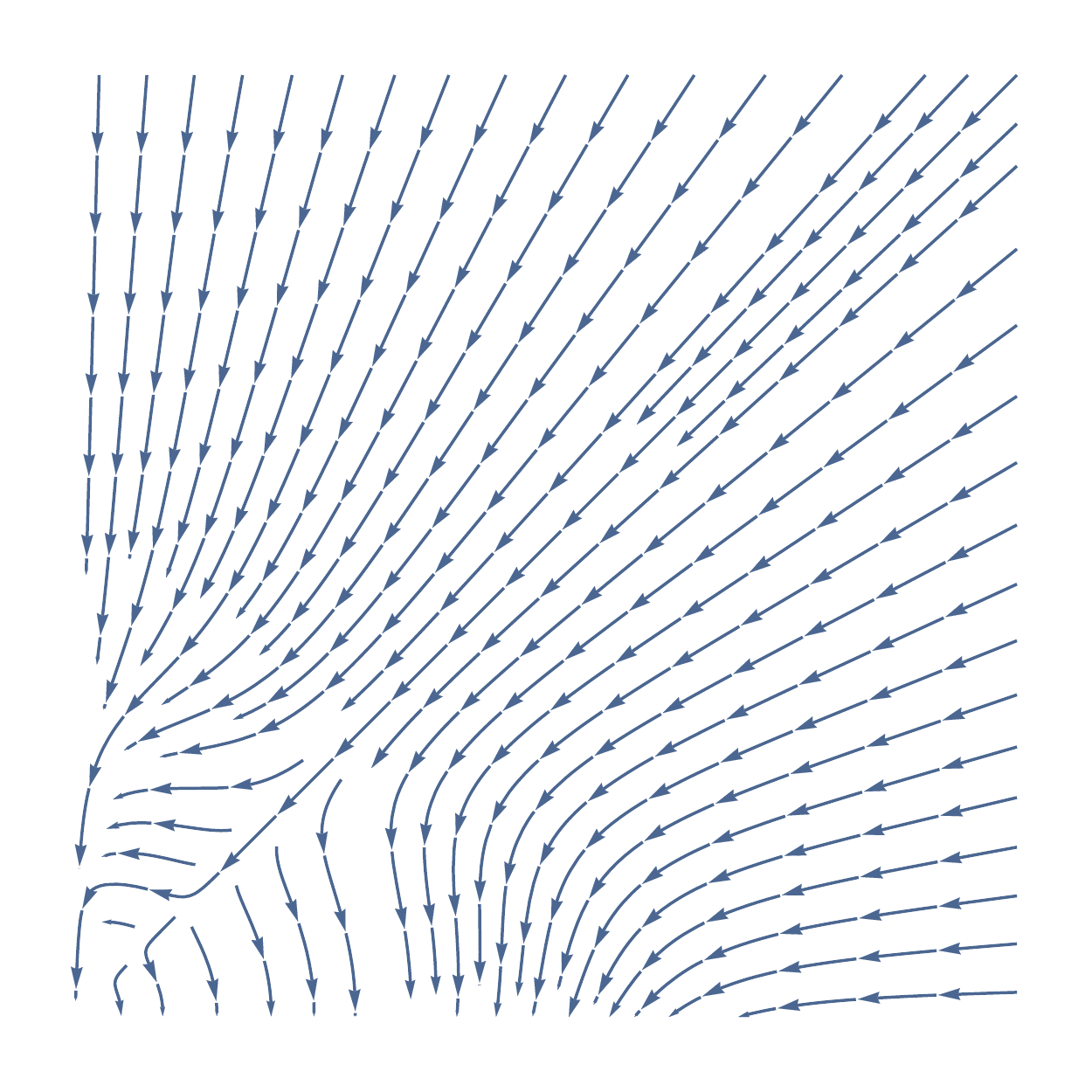}
		\caption{For $N=3$}
		\label{fig:N=3_psi=40_earlier}
	\end{subfigure}
	\begin{subfigure}{.5\textwidth}
		\centering
		\includegraphics[width=.8\linewidth]{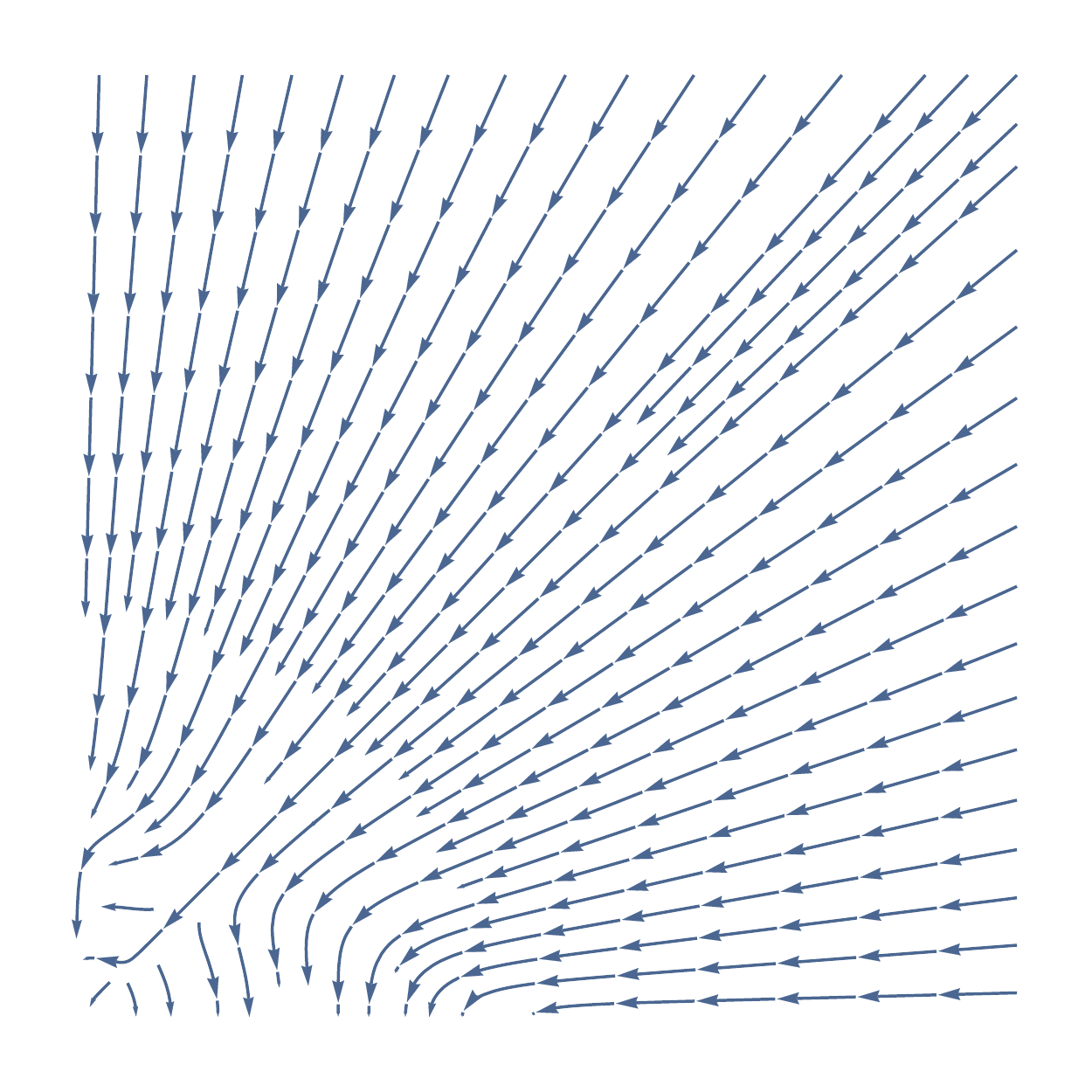}
		\caption{For $N=4$}
		\label{fig:for_N=4_psi=40_earlier}
	\end{subfigure}%
	\begin{subfigure}{.5\textwidth}
		\centering
		\includegraphics[width=.8\linewidth]{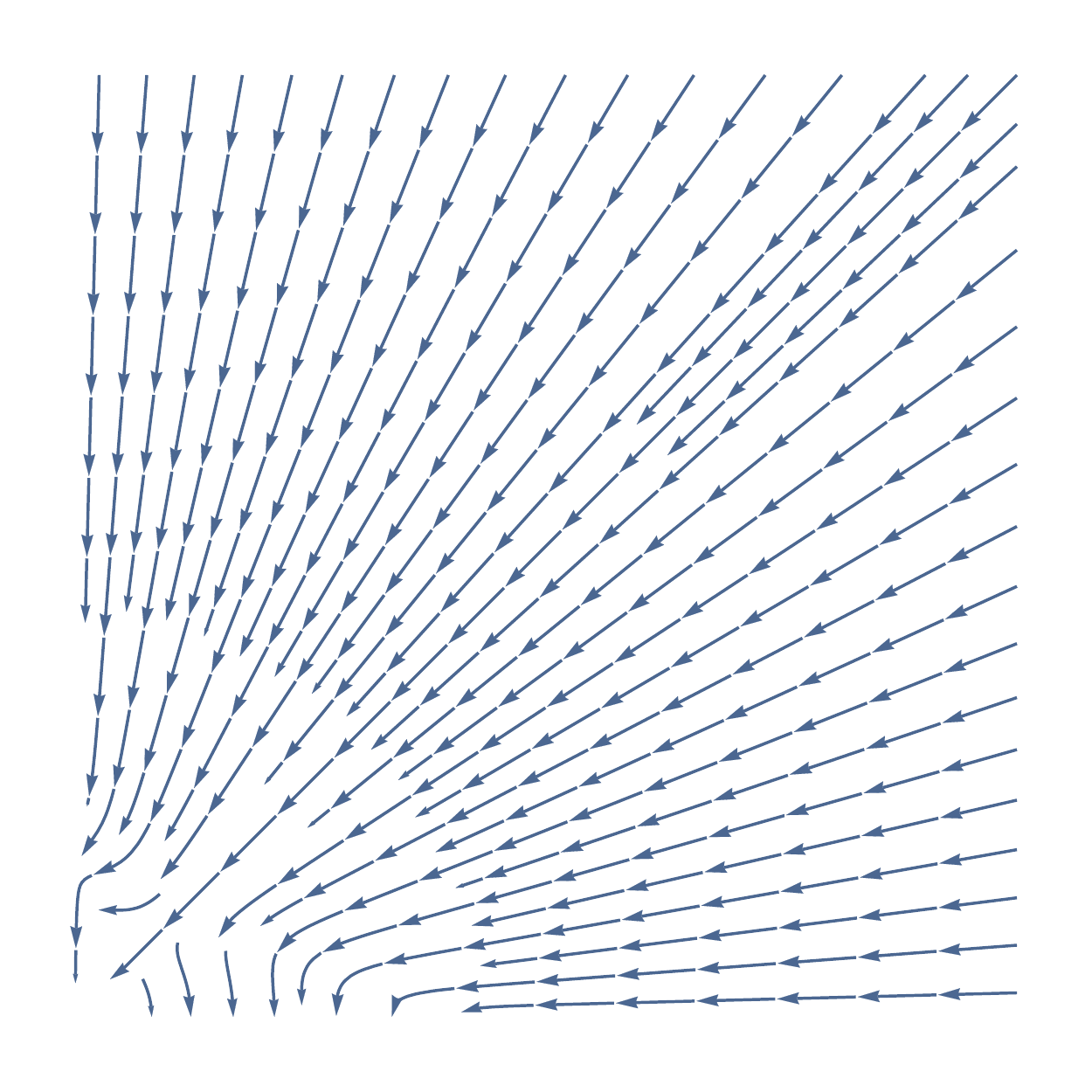}
		\caption{For $N=5$}
		\label{fig:for_N=5_psi=40_earlier}
	\end{subfigure}
	\caption{Earlier stage evolution of the fields $\theta$ and $\psi$  to the minimum of the potential of Eq.(\ref{toy_potential_1_exact_form}) for $A=B$, in the quadrant $0<\theta<\frac{\pi}{2}$. Here the flow starts at nearly around $\psi=40$}
	\label{fig:Early_stage_evolution_of_fields}
\end{figure}
\begin{figure}
	\centering
	\begin{subfigure}{.5\textwidth}
		\centering
		\includegraphics[width=.8\linewidth]{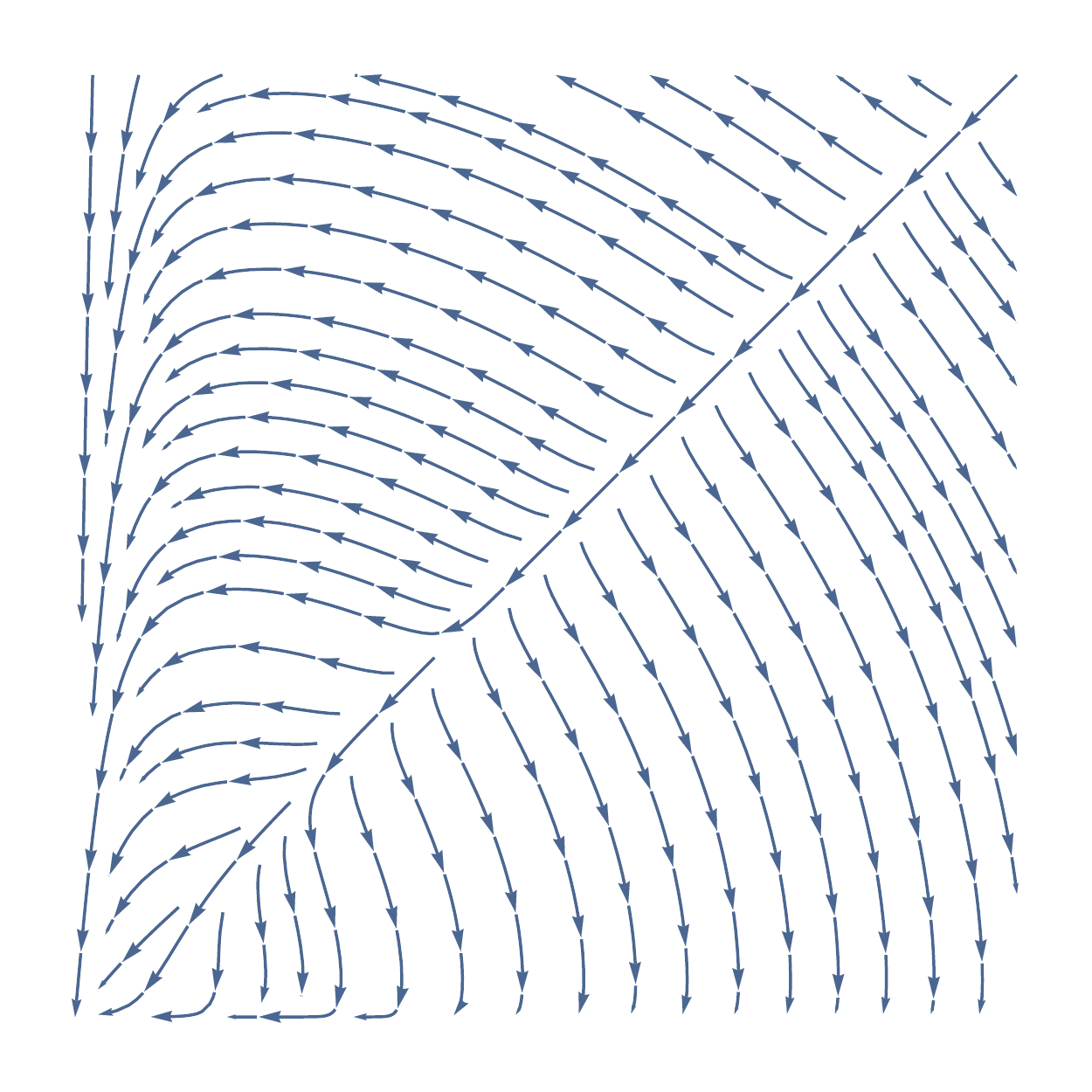}
		\caption{For $N=2$}
		\label{fig:for_N=2_psi=10_later}
	\end{subfigure}%
	\begin{subfigure}{.5\textwidth}
		\centering
		\includegraphics[width=.8\linewidth]{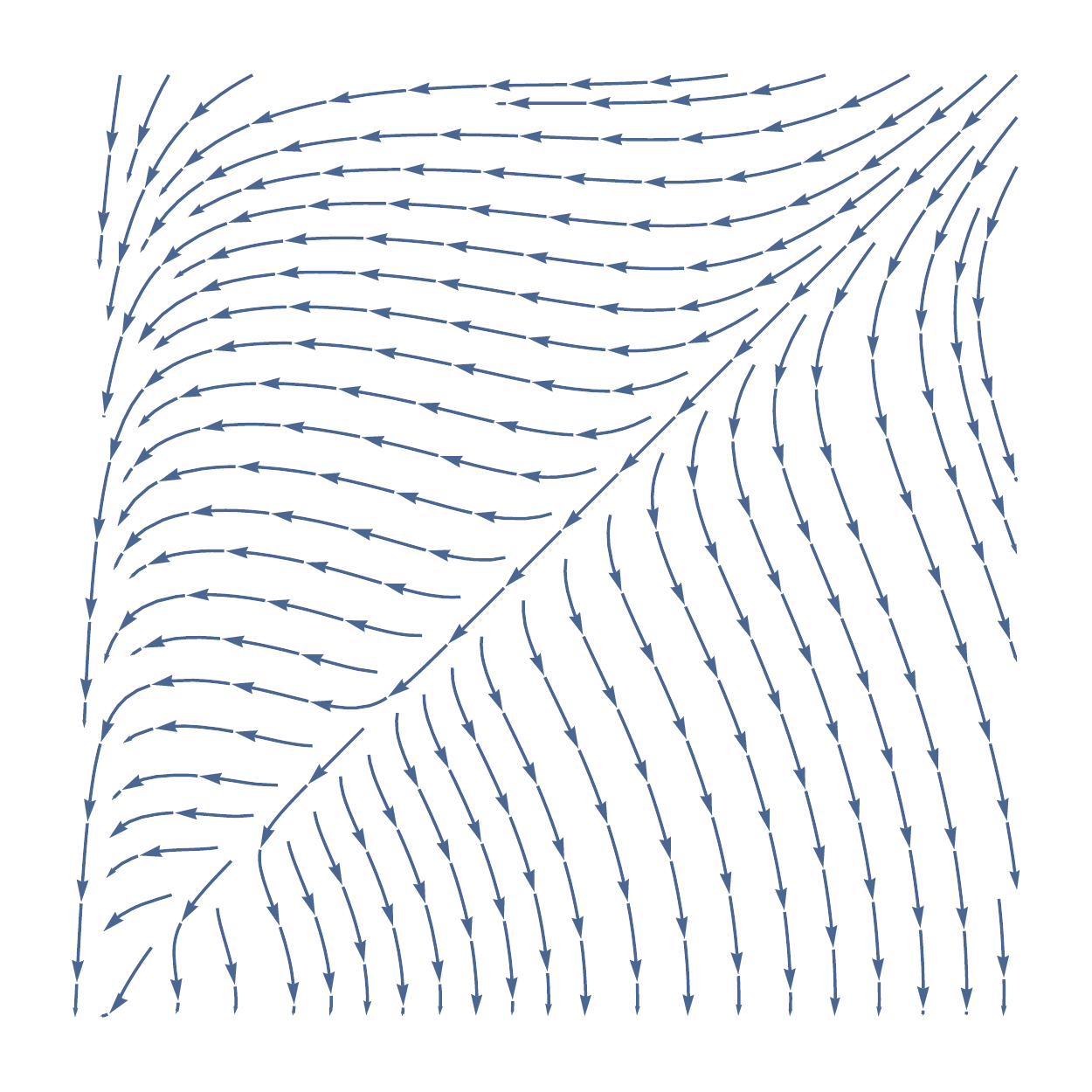}
		\caption{For $N=3$}
		\label{fig:for_N=3_psi=10_later}
	\end{subfigure}
	\begin{subfigure}{.5\textwidth}
		\centering
		\includegraphics[width=.8\linewidth]{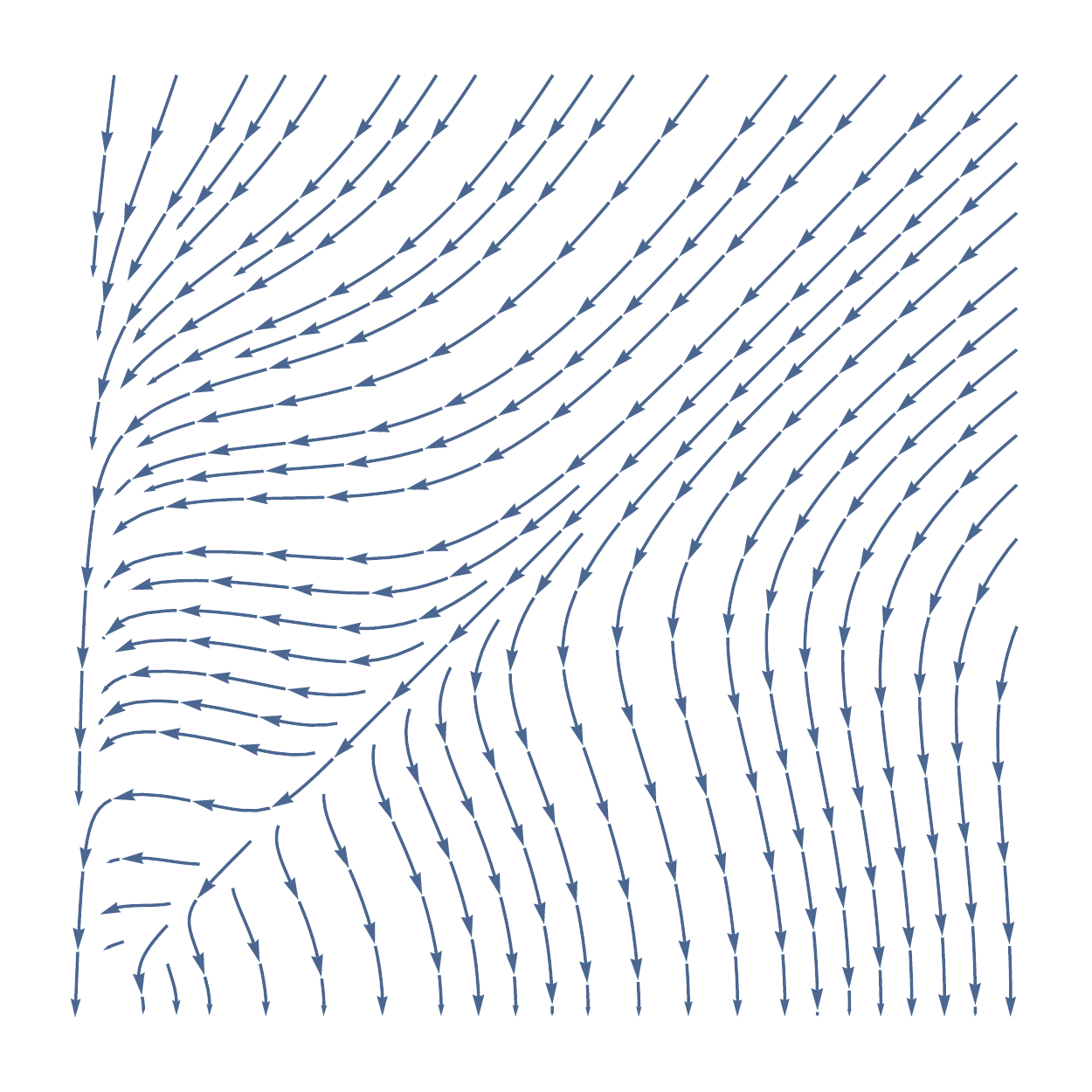}
		\caption{For $N=4$}
		\label{fig:for_N=4_psi=10_later}
	\end{subfigure}%
	\begin{subfigure}{.5\textwidth}
		\centering
		\includegraphics[width=.8\linewidth]{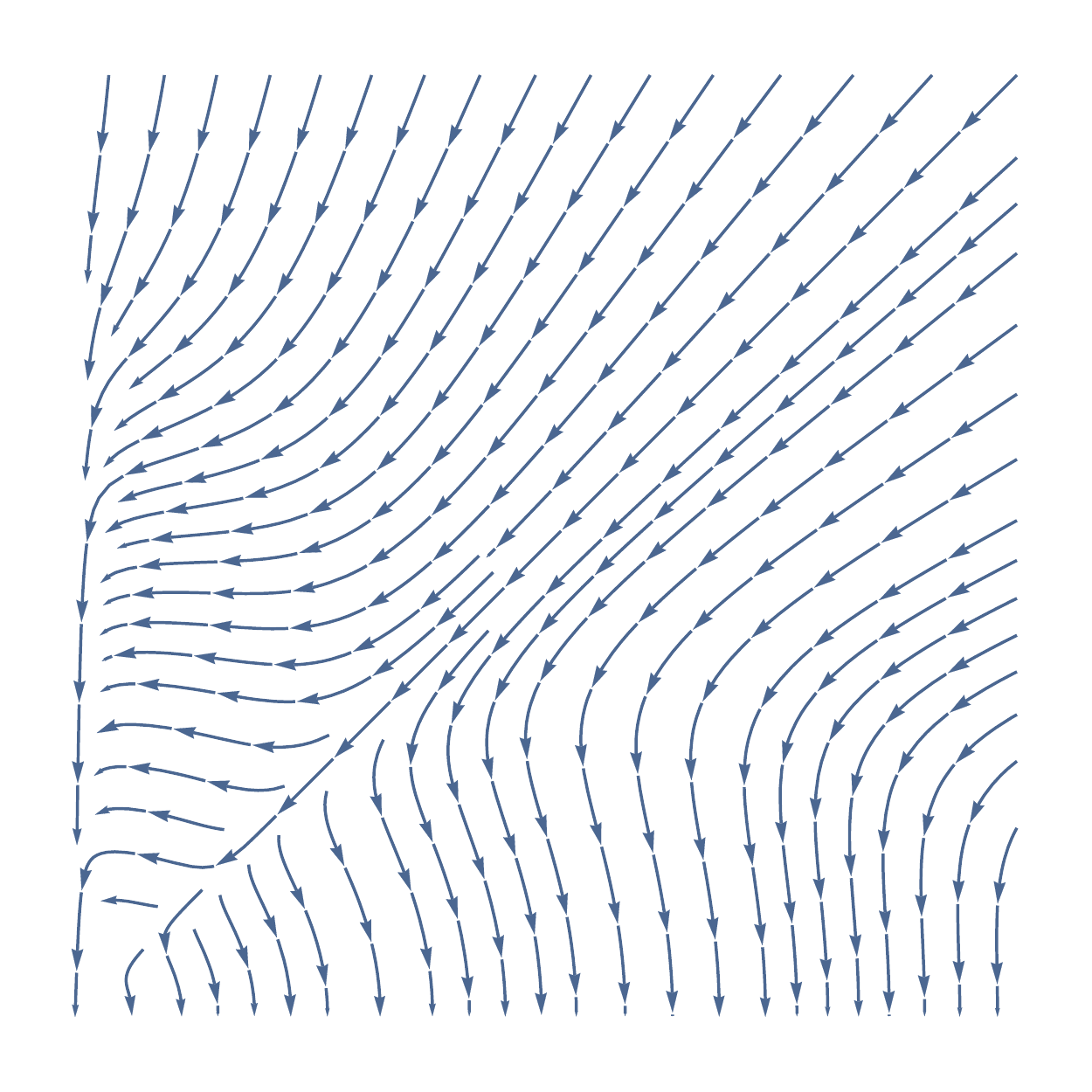}
		\caption{For $N=5$}
		\label{fig:for_N=5_psi=10_later}
	\end{subfigure}
	\caption{Flow of the fields $\psi$ and $\theta$ to the minimum of the potential Eq.(\ref{toy_potential_1_exact_form}) for $A=B$ at late stage of inflation in the quadrant $0<\theta<\frac{\pi}{2}$, flow starts at nearly around $\psi=10$.}
	\label{fig:Later_stage_evolution_of_fields}
\end{figure}
Consequently the radial derivative and angular derivative for the potential Eq.(\ref{toy_potential_approximated_for_large_psi}) can be expressed as follows:

\begin{equation}\label{radial_derivative_of_potential}
V_{\psi}=\left(A+2B\sin^{2}2\theta\right)\sqrt{\frac{2}{3}}4e^{-\sqrt{\frac{2}{3}}\psi}
\end{equation}
and
\begin{equation}\label{angular_derivative_of_potential}
V_{\theta}=2B\sin4\theta\left[\left(N-1\right)-8e^{-\sqrt{\frac{2}{3}}\psi}\right]
\end{equation}
Eq.(\ref{radial_derivative_of_potential}) clearly demonstrates that the radial derivative of the potential is exponentially suppressed, which means that for large $\psi$ regime potential is exponentially stretched to be very flat in the radial direction, and one can notice that this exponentially flattening of the potential will not be disturbed with respect to the consistent modification of the potential, i.e., in both sense, with respect to the consistent modification in the value of $N$ and in modification of any power of the potential. Also from Eq.(\ref{angular_derivative_of_potential}) the suppression in the angular derivative is negligible compared to the other term and as an output one can expect ridges and valleys in the angular direction due to the presents of trignometric sine function. Using Eq.(\ref{angular_derivative_of_potential}) in Eq.(\ref{very_approximated_eq_motion for_theta_field}), the velocity of angular field $\theta$ can be expressed as
\begin{equation}\label{final_speed_of_theta_field}
\dot{\theta}\simeq\frac{-2B\sin4\theta\left[\left(N-1\right)-8e^{-\sqrt{\frac{2}{3}}\psi}\right]}{3H\sum_{h=2}^{N}\left(A^{(h-4)}(\theta)-\frac{\left[A^{(h-2)}(\theta) \right ]^{2}}{A^{(h)}(\theta)} \right )\frac{(\sqrt{6})^{h}}{4}\frac{\sin^{2}2\theta }{2^{h}}e^{\frac{h\psi}{\sqrt{6}}}\left(1-he^{-\sqrt{\frac{2}{3}}\psi} \right )}
\end{equation} 
Also velocity of $\psi$ field can be estimated using Eq.(\ref{final_speed_of_theta_field}) and Eq.(\ref{toy_potential_approximated_for_large_psi}) in Eq.(\ref{very_approximated_eq_motion for_psi_field}) as:
\begin{multline}\label{final_speed_of_psi_field}
\dot{\psi}\simeq\frac{\left(A+2B\sin^{2}2\theta\right)\sqrt{\frac{2}{3}}4e^{-\sqrt{\frac{2}{3}}\psi}}{3H\left\{1+\left[ \frac{N(N+1)(2N+1)-30}{24}\right ]\left(1+4e^{-\sqrt{\frac{2}{3}}\psi} \right )\right\}}\\+\frac{\sum_{h=2}^{N}\left(A^{(h-4)}(\theta)-\frac{\left[A^{(h-2)}(\theta) \right ]^{2}}{A^{(h)}(\theta)} \right )\frac{\left(\sqrt{6}\right)^{h-1}}{2^{h}.8}h\sin^{2}2\theta\left(1+(2-h)e^{-\sqrt{\frac{2}{3}}\psi} \right )e^{\frac{h}{\sqrt{6}}\psi}}{3H\left\{1+\left[ \frac{N(N+1)(2N+1)-30}{24}\right ]\left(1+4e^{-\sqrt{\frac{2}{3}}\psi} \right )\right\}}\\ \times \frac{4B^{2}\sin^{2}4\theta\left[\left(N-1\right)-8e^{-\sqrt{\frac{2}{3}}\psi}\right]^{2}}{3\left[\sum_{h=2}^{N}\left(A^{(h-4)}(\theta)-\frac{\left[A^{(h-2)}(\theta) \right ]^{2}}{A^{(h)}(\theta)} \right )\frac{\left(\sqrt{6}\right)^{h}}{2^{h}.4}\sin^{2}2\theta\left(1-he^{-\sqrt{\frac{2}{3}}\psi} \right )e^{\frac{h}{\sqrt{6}}\psi}\right]^{2}}\\ \times \frac{1}{A\left[\left(N-1\right)-4e^{-\sqrt{\frac{2}{3}}\psi}\right]+B\sin^{2}2\theta\left[\left(N-1\right)-8e^{-\sqrt{\frac{2}{3}}\psi}\right]} 
\end{multline}
In order to understand the dynamics better, these velocities Eq.(\ref{final_speed_of_theta_field}) and Eq.(\ref{final_speed_of_psi_field}) can further be approximated to; 
\begin{equation}\label{final_very_approx_speed_of theta_field}
\dot{\theta}\simeq\frac{-2B\left(N-1\right)\sin4\theta}{3H\sum_{h=2}^{N}\left(A^{(h-4)}(\theta)-\frac{\left[A^{(h-2)}(\theta) \right ]^{2}}{A^{(h)}(\theta)} \right )\frac{(\sqrt{6})^{h}}{4}\frac{\sin^{2}2\theta }{2^{h}}e^{\frac{h\psi}{\sqrt{6}}}}
\end{equation} 
\begin{equation}\label{final_approx_speed_for_psi_field}
\dot{\psi}\simeq\frac{\left(A+2B\sin^{2}\theta\right)\sqrt{\frac{2}{3}}4e^{-\sqrt{\frac{2}{3}}\psi}}{3H\left\{1+\left[ \frac{N(N+1)(2N+1)-30}{24}\right ]\right\}}
\end{equation}
In this approximation one can clearly see that the contributions arising from the kinetic mixing terms are absent and can conclude that kinetic mixing has no role in the dynamics of inflation. Also from these equations one can observe that the contribution to numerator parts are arising from potential term and the contribution to denominator part is arising from kinetic terms. 
Thus 
the inflationary dynamics are stable not only with respect to the significant modifications of the radial potential, but also with respect to the significant modification of the angular potential. Thus our model shows a double attractor behavior. This is because, the value of $N$ arising from these potentials have negligibly small role in the dynamics of these fields compared to large value of $\psi$ field. But the dynamics will be hugely disturbed with the significant modification in the geometry of these fields, which means with respect to the significant modification in the non-canonical sector of the kinetic terms. In order to get more insight into these dynamics, from Equations Eq.(\ref{final_very_approx_speed_of theta_field}) and Eq.(\ref{final_approx_speed_for_psi_field}) one can readily check that for $N=2$, the field velocities are suppressed by the same factor $e^{-\sqrt{\frac{2}{3}}\psi}$. Since the evolution of $\psi$ field start at large values and range of evolution of the angular field is much small compared to the radial field, i.e., $\Delta\theta=O(1)$, field trajectory will be almost orthogonal to the ridge and as result $\theta$ field quickly rolls down to the valleys of the potential, i.e., to the valley at $\theta=0$ and at $\theta=\frac{\pi}{2}$ from the ridge at $\theta=\frac{\pi}{4}$, see the figures Fig.(\ref{fig:for_N=2_psi=40_earlier}) and Fig.(\ref{fig:for_N=2_psi=10_later}). Thus multi-field effect in this case quickly boils down to the single field scenario in the early stage of inflation and it is known as multi-field conformal attractors \cite{kallosh2013multi}. Now if go for the values of $N>2$, i.e., if we consider the modification in the geometry of the field space metric, speed of angular field will be more suppressed than the speed of radial field. This is evident from the equations Eq.(\ref{final_very_approx_speed_of theta_field}) and Eq.(\ref{final_approx_speed_for_psi_field}). As a result one can see that net combined evolution of these fields will be along the radial direction and will start to roll on the ridge of the potential. When the value of the $N$ becomes higher and higher, suppression in the speed of angular field increases and rolling on the ridge will occur for the long duration, this can be seen from figures Fig(\ref{fig:N=3_psi=40_earlier}), Fig(\ref{fig:for_N=4_psi=40_earlier}), and Fig(\ref{fig:for_N=5_psi=40_earlier}). Thus non-trivial multi-field effects come into play in this class of models unlike in the case of multi-field conformal attractors. So we have to calculate the multi-field inflationary parameters and their predictions for our model and make sure that this is compatible with the latest Planck data. This is the main difference between our model and the model described in \cite{kallosh2013multi}. A similar kind of dynamics, that is rolling on the ridge, can be observed in the model $\alpha$-attractors with $\alpha=1/3$ \cite{ana_achucarro_multi-field_alph_attra}. The main difference between this model and our model is coming from its construction. That is in $\alpha$-attractors with $\alpha=1/3$ the fields $\psi$ and $\theta$ are appears as partners, i.e., they are originating from the same super field $T$. Both of them are light fields and no fields are stabilized in \cite{ana_achucarro_multi-field_alph_attra}. But in our model these two fields are arising from two different superfields and their partners are stabilized during inflation.

\section{Inflationary parameters and predictions}\label{inflationary_parameters}
Here we use the $\delta N$-formalism \cite{m.sasaki_deltaN_formalism,lyth_non-gaussianity_from_deltaN_formalism} to compute the inflationary parameters and their predictions of our model defined in Eq.(\ref{eq_multifield_final_lagrangian}). For this we start by representing the following potential
\begin{equation}
V=\sum_{h=2}^{N}F\left (\tanh ^{\frac{2}{h}}\left [ \sinh^{-1}\left[\frac{\left(\sqrt{6}\right)^{\frac{h}{2}}}{h}
\sqrt{\frac{2A^{(h)}(\theta)}{3}}\sinh^{\frac{h}{2}}
\left(\frac{\psi}{\sqrt{6}} \right ) 
\right ] \right ],\theta\right )
\end{equation}
in more general form as
\begin{equation}
V=\sum_{h=2}^{N}F\left(z(\psi),\theta\right)
\end{equation}
where 
\begin{equation}
z(\psi)=\tanh ^{\frac{2}{h}}\left [ \sinh^{-1}\left[\frac{\left(\sqrt{6}\right)^{\frac{h}{2}}}{h}
\sqrt{\frac{2A^{(h)}(\theta)}{3}}\sinh^{\frac{h}{2}}
\left(\frac{\psi}{\sqrt{6}} \right ) 
\right ]\right]
\end{equation}
For very large values of $\psi>>1$ one can compute the radial derivative of the potential as follows,
\begin{equation}
V_{\psi}=V_{z}2\sqrt{\frac{2}{3}}e^{-\sqrt{\frac{2}{3}}\psi}
\end{equation}
here $V_{z}=\frac{\partial \sum_{h=2}^{N}F}{\partial z}$ and the slow-roll equation of motion of the $\psi$-field
\begin{equation}
\left\{1+\left[ \frac{N(N+1)(2N+1)-30}{24}\right ]\left(1+4e^{-\sqrt{\frac{2}{3}}\psi} \right )\right\} 3H\dot{\psi}=-V_{\psi}
\end{equation}
can be represented in terms of e-folding number $N_{e}$, as
\begin{equation}\label{diff_equation}
\frac{d\psi}{dN_{e}}=\frac{V_{z}}{V}\frac{2}{\alpha}\sqrt{\frac{2}{3}}e^{-\sqrt{\frac{2}{3}}\psi}
\end{equation}
where we used $dN_{e}=-Hdt$, and $\alpha=1+\beta$, and we define $\beta$ as the amount of non-canonicity in the kinetic term as follows:
\begin{equation}\label{beta_equation}
\frac{N(N+1)(2N+1)-30}{24}=\beta
\end{equation}
Now by solving the differential equation Eq.(\ref{diff_equation}), we can write number of e-foldings $N_{e}$ in terms of the fields $\psi$ and $\theta$. During this integration one can keep $V_{z}/V$ nearly a constant for the large value of $\psi$, since $z\rightarrow1$ when $\psi\rightarrow\infty$. Thus number of e-foldings can be written as

\begin{equation}\label{e-foldings}
N_{e}=\frac{1}{D}e^{\sqrt{\frac{2}{3}}\psi}+C(\theta)
\end{equation}
where
\begin{equation}\label{D_equation}
D=\frac{V_{z}}{V}\frac{4}{3\alpha}
\end{equation}
and $C(\theta)$ is the integration constant which is in the order of 1 and it can be calculated by using the boundary conditions of inflation. With the use of Eq.(\ref{e-foldings}) one can compute the curvature perturbations $\zeta$ at the end of the inflation, by using the formula \cite{m.sasaki_deltaN_formalism},

\begin{equation}
\zeta=\delta N_{e}=\frac{\partial N_{e}}{\partial \psi}\delta\psi+\frac{\partial N_{e}}{\partial \theta}\delta\theta
\end{equation}
Using Eq.(\ref{e-foldings}) this can be written as
\begin{equation}\label{curvature_perturbations}
\zeta=\delta N_{e}=\frac{1}{D}\sqrt{\frac{2}{3}}e^{\sqrt{\frac{2}{3}}\psi}\delta\psi+\left(C_{\theta}-\frac{D_{\theta}}{D^{2}}e^{\sqrt{\frac{2}{3}}\psi}\right)\delta\theta
\end{equation}
Now from the Lagrangian Eq.(\ref{eq_multifield_final_lagrangian}) one can calcualate the fluctuations of the angular field $\theta$, in comparison with the radial field $\psi$ as 

\begin{equation}\label{field_fluactuations}
\delta\theta\simeq\frac{\alpha^{\frac{1}{2}}\left(1+\frac{4\beta}{\alpha}e^{-\sqrt{\frac{2}{3}}\psi}\right)^{\frac{1}{2}}\delta\psi}{\sin2\theta\sqrt{\sum_{h=2}^{N}\frac{\left(\sqrt{6}\right)^{h}}{2^{h}.4}\left(A^{(h-4)}(\theta)-\frac{\left[A^{(h-2)}(\theta) \right ]^{2}}{A^{(h)}(\theta)} \right )\left(1-he^{-\sqrt{\frac{2}{3}}\psi} \right ) e^{\frac{h}{\sqrt{6}}\psi}}}
\end{equation}
this equation Eq.(\ref{field_fluactuations}) reveals that fluctuations in angular field $\delta\theta$ is exponentially suppressed in comparison with the fluctuations of radial field $\delta\psi$. So the second term in the eq.(\ref{curvature_perturbations}) can be neglected. Also the slow-roll condition for non-canonical field $\psi$ can be calculated as
\begin{equation}
\epsilon_{\psi}=\left(\frac{V_{\psi}}{V}\right)^{2}\frac{1}{2(1+\alpha)}\left(1-\frac{4\alpha e^{-\sqrt{\frac{2}{3}}\psi}}{1+\alpha}\right)
\end{equation}
using Eq.(\ref{D_equation}) this can be further simplified to
\begin{equation}\label{epsilon_in_D}
\epsilon_{\psi}=\frac{3\alpha D^{2}}{4}e^{-2\sqrt{\frac{2}{3}}\psi}
\end{equation}
Thus the final form of curvature perturbation in terms of this slow-roll variable can be specify to
\begin{equation}
\zeta\simeq\sqrt{\frac{\alpha}{2\epsilon_{\psi}}}\delta\psi
\end{equation}
Finally power spectrum for curvature perturbations can be calculated using the $\delta N$ formula \cite{m.sasaki_deltaN_formalism}:
\begin{equation}
P_{\zeta}=\left(\frac{H}{2\pi}\right)^{2}h^{ab}\frac{\partial N_{e}}{\partial \phi^{a}}\frac{\partial N_{e}}{\partial \phi^{b}}
\end{equation}
For our model Eq.(\ref{eq_multifield_final_lagrangian}), this curvature power spectrum reads the form

\begin{equation}
P_{\zeta}=\left(\frac{H}{2\pi}\right)^{2}\left[\frac{2}{3\alpha D^{2}}e^{2\sqrt{\frac{2}{3}}\psi}\left(1-\frac{4\beta }{\alpha}e^{-\sqrt{\frac{2}{3}}\psi}\right)\right]
\end{equation}
In terms of slow-roll parameter this further reduces to

\begin{equation}
P_{\zeta}=\frac{H^{2}}{8\pi^{2}\epsilon_{\psi}}\left(1-\frac{4\beta }{\alpha}e^{-\sqrt{\frac{2}{3}}}\psi\right)
\end{equation}
Moreover the slow-roll parameters can be expressed in terms of the number of e-foldings using Eq.(\ref{e-foldings}) and Eq.(\ref{epsilon_in_D}), as
\begin{equation}
\epsilon_{\psi}\simeq\frac{3\alpha}{4N_{e}^{2}},~~~~~~~~~\eta\simeq-\frac{1}{N_{e}}
\end{equation}
From the above, it can be seen that, this slow-roll parameter expression is same as that of the expression of the slow-roll parameter of single field non-canonical conformal attractors \cite{pinhero2017non-cano-con-attra-sing-inf}, if we neglect the extra correction term $\frac{\sqrt{3\beta}}{2}$ compared to the large value of $N_e$ in \cite{pinhero2017non-cano-con-attra-sing-inf}. As a result the inflationary parameter predictions for this multi-field non-canonical conformal attractors  boils down to the predictions of single field non-canonical conformal attractors;
\begin{equation}
1-n_{s}\simeq\frac{2}{N_{e}},~~~~~~~~~~~~r\simeq\frac{12\alpha}{N_{e}}
\end{equation}
 where $\alpha$ can take only values $\alpha=1,3.25,7.25,\dots$ for $N=2,3,4,\dots$ respectively, which is evident from Eq.(\ref{beta_equation}).

Another noteworthy aspect of multi-field inflationary models is that it will predict huge local non-Guassianity $f_{NL}$. Planck 2015 and Planck 2018 data disfavors this large non-Gaussian features. So it is important to check if our model can pass this tight constraint provided by Planck. Here we calculate the size of the bispectrum from our model using the method adopted by \cite{lyth_non-gaussianity_from_deltaN_formalism}. Expanding $\delta N$ formula upto the second order field fluctuations as,
\begin{equation}\label{bispectrum}
\zeta=\delta N_{e}=\frac{\partial N_{e}}{\partial \psi}\delta\psi+\frac{\partial N_{e}}{\partial \theta}\delta\theta+\frac{1}{2}\frac{\partial^{2}N_{e}}{\partial\psi^{2}}\delta\psi^{2}+\frac{1}{2}\frac{\partial^{2}N_{e}}{\partial\theta^{2}}\delta\theta^{2}+\frac{\partial^{2}N_{e}}{\partial\psi\partial\theta}\delta\psi\delta\theta
\end{equation}
By considering the field fluctuations in $\theta$ direction is exponentially suppressed as shown in Eq.(\ref{field_fluactuations}),  and also by assuming there is no large coupling between the field fluctuations, one can conclude that, only first and third term in Eq.(\ref{bispectrum}) will contribute to the bispectrum. As a result, the local non-Gaussianity can be approximately computed from Eq.(\ref{bispectrum}) as;
\begin{equation}
\frac{5}{6}f_{NL}=\frac{\partial^{2}N_{e}}{\partial\psi^{2}}\bigg/ \left(\frac{\partial N_{e}}{\partial \psi}\right)^{2}\simeq\frac{5}{6N_{e}}
\end{equation}
Again this will boils down to the prediction of single field inflation model provided by the consistency relation $f_{NL}=-\frac{5}{12}(n_{s}-1)$ \cite{maldacena_non-gaussian_consistancy_relation}. Thus our model represents multi-field conformal inflation with non-canonical kinetic terms in both radial and angular fields with multi-field effects in its dynamics but with predictions identical to single field non-canonical conformal inflation model \cite{pinhero2017non-cano-con-attra-sing-inf}. This is an unique feature of our model in comparison to the other multi-field models available in the literature (except $\alpha$-attractors with $\alpha=1/3$ model \cite{ana_achucarro_multi-field_alph_attra}) in the sense that it reflects true multi-field dynamics that eventually leads to single field predictions.



\section{Summary and Conclusions}\label{summary_and_conclusions}
In this paper we developed a multi-field generalization of recently discovered class of non-canonical conformal attractors \cite{pinhero2017non-cano-con-attra-sing-inf}, which also generalized the idea of multi-field conformal attractors \cite{kallosh2013multi} simultaneously. For this we consider a Lagrangian in $\mathcal{N}=1$ superconformal theory with two conformon fields and multiple inflaton fields, which are non-canonical in their kinetic forms.  Conformal breaking of this theory can produce a class of non-canonical models, which can govern multi-field inflation compatible with the recent observations. In this generalization we found that multi-field effects are arising when the amount of non-canonicity increases in the kinetic term. In conformal multi-field scenario, inflaton is first rolling down to the valley from the ridge and then move to minimum of the potential along the valley, so in the early stage of inflation and onwards, model is behaving as the single field model. But in our model due to the effect of non-canonical terms in the original conformal variables, speed of angular field is hugely suppressed. Hence, instead of first rolling down to the valley, inflaton starts to roll on the ridge,  this is shown in the Figures, Fig.(\ref{fig:Early_stage_evolution_of_fields}) and Fig.(\ref{fig:Later_stage_evolution_of_fields}). Then we use the $\delta N$ formalism for the calculations of inflationary parameters. We found that, even though multi-field effects and dynamics are there in our model, inflationary predictions are boil down to the predictions of single field non-canonical conformal attractors case. This is because, curvature fluctuations in the $\theta$ direction is exponentially suppressed, due to presence of non-canonical kinetic terms in the theory. Approximately  same dynamics is also observed in $\alpha$-attractors with $\alpha=1/3$ models \cite{ana_achucarro_multi-field_alph_attra}. But the inflationary prediction for $r$ value are slightly differed with respect to our model. Thus one can also say our model is partially mimicking the dynamics of  $\alpha$-attractors with $\alpha=1/3$. We further show that our model is quite consistent with latest Planck 2018 data.

In conclusion, our main observation is that, inflationary predictions and dynamics in these class of models are stable with respect to the strong modifications of the potential. i.e., our model shows a double attractor behaviour, since the cosmological predictions are not altered, with respect to the strong modifications of both radial and angular part of potential. 
 

\section*{Acknowledgments}

TP thanks  A. Chatterjee, D. Chandra, A.Naskar and A. Paul
for their fruitful suggestions and discussions. 
TP is supported by Senior Research fellowship 
(Order No. DS/18-19/616) of the Indian Statistical 
Institute (ISI), Kolkata.
\begin{appendices}
	\section{Explicit analysis on the evolution of fields during inflation}\label{appendix_1}
	For the  the background dynamics, one can calculate the equation of motion for scalar fields $\psi$ and $\theta$, from the action Eq.(\ref{eq_multifield_final_lagrangian}), for the FLRW spacetime metric, as follows:
	\begin{multline}\label{equ_of_motion_for_psi_field}
	\sum_{h=2}^{N}\left \{\frac{\left(\sqrt{6}\right)^{h-2}A^{(h)}(\theta)
		\sinh^{h-2}\left(\frac{\psi}{\sqrt{6}}\right)
		\cosh^{2}\left(\frac{\psi}{\sqrt{6}}\right)}{\left[1+
		\frac{2\left(\sqrt{6}\right)^{h}}{3h^{2}}A^{(h)}(\theta)\sinh^{h}
		(\frac{\psi}{\sqrt{6}})\right]}\left ( 3H\dot{\psi}+\ddot{\psi} \right )\right. \\  \left.+\left[ \left(1+\frac{2\left(\sqrt{6}\right)^{h}}{3h^{2}}A^{(h)}(\theta)\sinh^{h}
	\left(\frac{\psi}{\sqrt{6}}\right)\right)\left((h-2)\sinh^{h-3}\left(\frac{\psi}{\sqrt{6}} \right )\cosh^{3}\left(\frac{\psi}{\sqrt{6}} \right )\right.\right. \right.\\
	\left. \left.\left.+\sinh^{h-2}\left(\frac{\psi}{\sqrt{6}} \right )\sinh\left(\frac{2\psi}{\sqrt{6}} \right ) \right )-\frac{2\left(\sqrt{6}\right)^{h}}{3h}A^{(h)}(\theta)\sinh^{2h-3}\left(\frac{\psi}{\sqrt{6}} \right )\cosh^{3}\left(\frac{\psi}{\sqrt{6}} \right )\right ]\right. \\  \left.\times\frac{\left(\sqrt{6}\right)^{h-1}A^{(h)}(\theta)}{12\left[1+\frac{2\left(\sqrt{6}\right)^{h}}{3h^{2}}A^{(h)}(\theta)\sinh^{h}
		\left(\frac{\psi}{\sqrt{6}}\right)\right]^{2}}\dot{\psi}^{2}-\frac{\left(\sqrt{6}\right)^{h-1}}{8}h\sin^{2}2\theta\sinh^{h-1}\left(\frac{\psi}{\sqrt{6}}\right)
	\right. \\  \left.\times\cosh\left(\frac{\psi}{\sqrt{6}}\right)\left[\frac{[A^{(h-2)}(\theta)]^{2}}{A^{(h)}(\theta)}-A^{(h-4)}(\theta)- \frac{[A^{(h-2)}(\theta)]^{2}}{A^{(h)}(\theta)\left[1+\frac{2\left(\sqrt{6}\right)^{h}}{3h^{2}}A^{(h)}(\theta)\sinh^{h}(\frac{\psi}{\sqrt{6}}) \right ]^{2}} \right ]\dot{\theta}^{2}\right. \\  \left.-\left[\frac{\left(\sqrt{6}\right)^{h}}{3h}[A^{(h-2)}(\theta)]^{2}\sin^{2}\theta\sinh^{2h-1}\left(\frac{\psi}{\sqrt{6}}\right)+\left(1+\frac{2\left(\sqrt{6}\right)^{h}}{3h^{2}}A^{(h)}(\theta)\sinh^{h}\left(\frac{\psi}{\sqrt{6}}\right) \right )\right.\right. \\
	\left. \left.\left(2A^{(h-2)}(\theta)\cos2\theta-\frac{(h-2)}{2}A^{(h-4)}(\theta)\sin^{2}2\theta \right )\sinh^{h-1}\left(\frac{\psi}{\sqrt{6}}\right)\right]\frac{\left(\sqrt{6}\right)^{h+1}\cosh\left(\frac{\psi}{\sqrt{6}}\right)}{12\left[1+\frac{2\left(\sqrt{6}\right)^{h}}{3h^{2}}A^{(h)}(\theta)\sinh^{h}\left(\frac{\psi}{\sqrt{6}}\right) \right ]^{2}}\dot{\theta}^{2}\right. \\  \left.-\frac{\left(\sqrt{6}\right)^{h-1}A^{(h-2)}(\theta)
		\sin2\theta\sinh^{h-1}\left(\frac{\psi}{\sqrt{6}}\right)
		\cosh\left(\frac{\psi}{\sqrt{6}}\right)}{2\left[1+
		\frac{2\left(\sqrt{6}\right)^{h}}{3h^{2}}A^{(h)}(\theta)\sinh^{h}
		(\frac{\psi}{\sqrt{6}})\right]}\left ( 3H\dot{\theta}+\ddot{\theta} \right )\right. \\  \left.-\frac{\left(\sqrt{6}\right)^{h}hA^{(h-2)}(\theta)\sin2\theta\sinh^{h-2}\left(\frac{\psi}{\sqrt{6}} \right )\cosh^{2}\left(\frac{\psi}{\sqrt{6}} \right )}{12\left[1+\frac{2\left(\sqrt{6}\right)^{h}}{3h^{2}}A^{(h)}(\theta)\sinh^{h}
		\left(\frac{\psi}{\sqrt{6}}\right)\right]^{2}}\dot{\psi}\dot{\theta}\right\}=-V_{\psi}
	\end{multline}
	\begin{multline}\label{equ_of_motion_for_theta_field}
	\sum_{h=2}^{N}\left \{-\frac{\left(\sqrt{6}\right)^{h-1}A^{(h-2)}(\theta)\sin2\theta
		\sinh^{h-1}\left(\frac{\psi}{\sqrt{6}}\right)
		\cosh\left(\frac{\psi}{\sqrt{6}}\right)}{2\left[1+
		\frac{2\left(\sqrt{6}\right)^{h}}{3h^{2}}A^{(h)}(\theta)\sinh^{h}
		(\frac{\psi}{\sqrt{6}})\right]}\left ( 3H\dot{\psi}+\ddot{\psi} \right )\right. \\  \left.-\left[h\sinh^{h-2}\left(\frac{\psi}{\sqrt{6}}\right)\cosh^{2}\left(\frac{\psi}{\sqrt{6}}\right)+\left(1+\frac{2\left(\sqrt{6}\right)^{h}}{3h^{2}}A^{(h)}(\theta)\sinh^{h}\left(\frac{\psi}{\sqrt{6}}\right) \right )\right.\right. \\  \left.\left.
	\times\left(\sinh^{h}\left(\frac{\psi}{\sqrt{6}}\right) -\sinh^{h-2}\left(\frac{\psi}{\sqrt{6}}\right)\cosh^{2}\left(\frac{\psi}{\sqrt{6}}\right)  \right )\right]\frac{\left(\sqrt{6}\right)^{h}A^{(h-2)}(\theta)\sin2\theta}{12\left[1+\frac{2\left(\sqrt{6}\right)^{h}}{3h^{2}}A^{(h)}(\theta)\sinh^{h}\left(\frac{\psi}{\sqrt{6}}\right) \right ]^{2}}\dot{\psi}^{2}\right. \\  \left.-\left[\frac{(\sqrt{6})^{h}}{4}\sin^{2}2\theta\sinh^{h}(\frac{\psi}{\sqrt{6}})\left(\frac{[A^{(h-2)}(\theta)]^{2}}{A^{(h)}(\theta)}-A^{(h-4)}(\theta) \right )-\frac{(\sqrt{6})^{h}[A^{(h-2)}(\theta)]^{2}\sin^{2}2\theta\sinh^{h}(\frac{\psi}{\sqrt{6}})}{4A^{(h)}(\theta)\left[1+\frac{2\left(\sqrt{6}\right)^{h}}{3h^{2}}A^{(h)}(\theta)\sinh^{h}(\frac{\psi}{\sqrt{6}}) \right ]} \right ]\right. \\  \left.\left(\ddot{\theta}+3H\dot{\theta} \right )+\left[\left(1+\frac{4(\sqrt{6})^{h}}{3h^{2}}A^{(h)}(\theta)\sinh^{h}(\frac{\psi}{\sqrt{6}}) \right )\frac{h}{2}[A^{(h-2)}(\theta)]^{3}\sin^{3}2\theta\right.\right. \\  \left.\left.+A^{(h)}(\theta)\left(1+\frac{2(\sqrt{6})^{h}}{3h^{2}}A^{(h)}(\theta)\sinh^{h}(\frac{\psi}{\sqrt{6}}) \right )\left(2[A^{(h-2)}(\theta)]^{2}\sin4\theta-A^{(h-2)}(\theta)A^{(h-4)}(\theta)(h-2)\sin^{3}2\theta \right )\right]\right. \\  \left.\times\frac{\left(\sqrt{6}\right)^{h}[A^{(h-2)}(\theta)]^{-2}\sinh^{h}(\frac{\psi}{\sqrt{6}})}{8\left[1+\frac{2\left(\sqrt{6}\right)^{h}}{3h^{2}}A^{(h)}(\theta)\sinh^{h}(\frac{\psi}{\sqrt{6}}) \right ]^{2}}\dot{\theta}^{2}-\left[2\sin4\theta\left(\frac{[A^{(h-2)}(\theta)]^{2}}{A^{(h)}(\theta)}-A^{(h-4)}(\theta) \right )\right.\right. \\  \left.\left.+\left(\frac{h[A^{(h-2)}(\theta)]^{2}}{[A^{(h)}(\theta)]^{2}}-\frac{2(h-2)A^{(h-2)}(\theta)A^{(h-4)}(\theta)}{A^{(h)}(\theta)}+(h-4)A^{(h-6)}(\theta) \right )\frac{\sin^{3}2\theta}{2} \right ]\frac{(\sqrt{6})^{h}\sinh^{h}(\frac{\psi}{\sqrt{6}}) }{8}\dot{\theta}^{2}\right. \\  \left.+\frac{\left(\sqrt{6}\right)^{h-1}h[A^{(h-2)}(\theta)]^{2}\sin^{2}2\theta\sinh^{h-1}(\frac{\psi}{\sqrt{6}})\cosh(\frac{\psi}{\sqrt{6}})}{4A^{(h)}(\theta)\left[1+\frac{2\left(\sqrt{6}\right)^{h}}{3h^{2}}A^{(h)}(\theta)\sinh^{h}(\frac{\psi}{\sqrt{6}}) \right ]^{2}}\dot{\psi}\dot{\theta}\right. \\  \left.-\left(\frac{[A^{(h-2)}(\theta)]^{2}}{A^{(h)}(\theta)}-A^{(h-4)}(\theta) \right )\frac{(\sqrt{6})^{h-1} }{4}\sin^{2}2\theta\sinh^{h-1}(\frac{\psi}{\sqrt{6}})\cosh(\frac{\psi}{\sqrt{6}})\dot{\psi}\dot{\theta}\right\}=-V_{\theta}
	\end{multline}
	Also by varying the action Eq.(\ref{eq_multifield_final_lagrangian}) with respect to the metric $g^{\mu\nu}$ we get the Friedmann's equation for the FLRW background as follows: 
	
	\begin{multline}\label{friedmann_equation}
	3H^{2}=\sum_{h=2}^{N}\left \{\frac{\left(\sqrt{6}\right)^{h}A^{(h)}(\theta)
		\sinh^{h-2}\left(\frac{\psi}{\sqrt{6}}\right)
		\cosh^{2}\left(\frac{\psi}{\sqrt{6}}\right)}{12\left[1+
		\frac{2\left(\sqrt{6}\right)^{h}}{3h^{2}}A^{(h)}(\theta)\sinh^{h}
		(\frac{\psi}{\sqrt{6}})\right]}\dot{\psi}^{2}\right. \\  \left.-\frac{\left(\sqrt{6}\right)^{h-1}A^{(h-2)}(\theta)\sin2\theta\sinh^{h-1}\left(\frac{\psi}{\sqrt{6}}\right)
		\cosh\left(\frac{\psi}{\sqrt{6}}\right)}{2\left[1+
		\frac{2\left(\sqrt{6}\right)^{h}}{3h^{2}}A^{(h)}(\theta)\sinh^{h}
		\left(\frac{\psi}{\sqrt{6}}\right)\right]}\dot{\psi}\dot{\theta}\right. \\  \left.-\left[\frac{\left[A^{(h-2)}(\theta) \right ]^{2}}{A^{(h)}(\theta)}-A^{(h-4)}(\theta) -\frac{\left[A^{(h-2)}(\theta)\right]^{2}
	}{A^{(h)}(\theta)\left[1+
		\frac{2\left(\sqrt{6}\right)^{h}}{3h^{2}}A^{(h)}(\theta)\sinh^{h}
		\left(\frac{\psi}{\sqrt{6}}\right)\right]}\right ]\right. \\  \left.\times\frac{\left(\sqrt{6}\right)^{h}}{8}\sin^{2}2\theta\sinh^{h}\left(\frac{\psi}{\sqrt{6}} \right )\dot{\theta}^{2}\right\}+V\left(\psi, \theta\right)
	\end{multline}
	As inflation takes place at very large values of the field, after quick period of relaxation, the fields can reach at slow-roll regime. So the slow-roll parameter for $\epsilon$ takes the form as,
	\begin{multline}
	\epsilon\equiv -\frac{\dot{H}}{H^{2}}=\frac{1}{H^{2}}\sum_{h=2}^{N}\left \{\frac{\left(\sqrt{6}\right)^{h}A^{(h)}(\theta)
		\sinh^{h-2}\left(\frac{\psi}{\sqrt{6}}\right)
		\cosh^{2}\left(\frac{\psi}{\sqrt{6}}\right)}{12\left[1+
		\frac{2\left(\sqrt{6}\right)^{h}}{3h^{2}}A^{(h)}(\theta)\sinh^{h}
		(\frac{\psi}{\sqrt{6}})\right]}\dot{\psi}^{2}\right. \\  \left.-\frac{\left(\sqrt{6}\right)^{h-1}A^{(h-2)}(\theta)\sin2\theta\sinh^{h-1}\left(\frac{\psi}{\sqrt{6}}\right)
		\cosh\left(\frac{\psi}{\sqrt{6}}\right)}{2\left[1+
		\frac{2\left(\sqrt{6}\right)^{h}}{3h^{2}}A^{(h)}(\theta)\sinh^{h}
		\left(\frac{\psi}{\sqrt{6}}\right)\right]}\dot{\psi}\dot{\theta}\right. \\  \left.-\left[\frac{\left[A^{(h-2)}(\theta) \right ]^{2}}{A^{(h)}(\theta)}-A^{(h-4)}(\theta) -\frac{\left[A^{(h-2)}(\theta)\right]^{2}
	}{A^{(h)}(\theta)\left[1+
		\frac{2\left(\sqrt{6}\right)^{h}}{3h^{2}}A^{(h)}(\theta)\sinh^{h}
		\left(\frac{\psi}{\sqrt{6}}\right)\right]}\right ]\right. \\  \left.\times\frac{\left(\sqrt{6}\right)^{h}}{8}\sin^{2}2\theta\sinh^{h}\left(\frac{\psi}{\sqrt{6}} \right )\dot{\theta}^{2} \right\}\ll 1
	\end{multline}
	From these slow-roll condition, one can readily see that non-canonical kinetic energies of these fields should be very small compared to the potential. As a result the Friedmann's equation defined in Eq.(\ref{friedmann_equation}) becomes
	\begin{equation}\label{appendix_approximated_friedmann_equation}
	3H^{2}\simeq V\left(\psi, \theta\right)
	\end{equation}
	As we are interested to study the evolution of fields during the inflation for asymptotically large values of $\psi$, we will represent all the hyperbolic functions in the equations Eq.(\ref{equ_of_motion_for_psi_field}) and Eq.(\ref{equ_of_motion_for_theta_field}) in terms of exponential functions and we keep the terms up to the first order corrections only. So in this approximation the equation of motion for scalar fields take the form 
	\begin{multline}\label{approximated_psi_equation}
	\left\{1+\left[ \frac{N(N+1)(2N+1)-30}{24}\right ]\left(1+4e^{-\sqrt{\frac{2}{3}}\psi} \right )\right\}\left ( 3H\dot{\psi}+\ddot{\psi} \right )-\left[N(N+1)(2N+1)-30\right] \\  \times\frac{e^{-\sqrt{\frac{2}{3}}\psi}}{6\sqrt{6}}\dot{\psi}^{2}-\frac{3\sin2\theta}{4\sqrt{6}}\left(1+2e^{-\sqrt{\frac{2}{3}}\psi} \right )\left(3H\dot{\theta}+\ddot{\theta} \right )\sum_{h=2}^{N}\frac{h^{2}A^{(h-2)}(\theta)}{A^{(h)}(\theta)}\\  -\sum_{h=2}^{N}\left[ \frac{[A^{(h-2)}(\theta)]^{2}h^{3}\sin^{2}2\theta}{[A^{(h)}(\theta)]^{2}}+\frac{h^{2}}{A^{(h)}(\theta)}\left[4A^{(h-2)}(\theta)\cos2\theta-(h-2)A^{(h-4)}(\theta)\sin^{2}2\theta \right ]\right ]\\  \times\frac{\sqrt{6}}{16}\left(1+2e^{-\sqrt{\frac{2}{3}}\psi} \right )\dot{\theta}^{2}+\sum_{h=2}^{N}\left(\frac{\left[A^{(h-2)}(\theta) \right ]^{2}}{A^{(h)}(\theta)}-A^{(h-4)}(\theta) \right )\frac{(\sqrt{6})^{h-1}}{8}\frac{h\sin^{2}2\theta }{2^{h}}e^{\frac{h\psi}{\sqrt{6}}}\\\times\left[1+(2-h)e^{-\sqrt{\frac{2}{3}}\psi} \right ]\dot{\theta}^{2}\simeq-V_{\psi}
	\end{multline}
	and 
	\begin{multline}\label{approximated_theta_equation}
	\sum_{h=2}^{N}\left\{-\frac{3\sin2\theta}{4\sqrt{6}}\left(1+2e^{-\sqrt{\frac{2}{3}}\psi} \right )\left(3H\dot{\psi}+\ddot{\psi} \right )\frac{h^{2}A^{(h-2)}(\theta)}{A^{(h)}(\theta)}+\frac{\sin2\theta}{2}e^{-\sqrt{\frac{2}{3}}\psi}\frac{h^{2}A^{(h-2)}(\theta)}{A^{(h)}(\theta)}\dot{\psi}^{2}\right. \\  \left.+\left[ \left(\frac{\left[A^{(h-2)}(\theta) \right ]^{2}}{A^{(h)}(\theta)}-A^{(h-4)}(\theta) \right )\frac{(\sqrt{6})^{h}}{4}\frac{\sin^{2}2\theta }{2^{h}}e^{\frac{h\psi}{\sqrt{6}}}\left(1-he^{-\sqrt{\frac{2}{3}}\psi} \right )-\frac{\left[A^{(h-2)}(\theta) \right ]^{2}}{[A^{(h)}(\theta)]^{2}}\frac{3h^{2}}{8}\sin^{2}2\theta \right ]\right. \\  \left.\times\left(3H\dot{\theta}+\ddot{\theta} \right )+\frac{3h^{2}}{16[A^{(h)}(\theta)]^{2}}\left(2\left[A^{(h-2)}(\theta) \right ]^{2}\sin4\theta -\frac{(h-2)}{2}A^{(h-2)}(\theta)A^{(h-4)}(\theta)\sin^{3}2\theta \right )\dot{\theta^{2}}\right. \\  \left.-\left[ \left(\frac{\left[A^{(h-2)}(\theta) \right ]^{2}}{A^{(h)}(\theta)}-A^{(h-4)}(\theta) \right )2\sin4\theta +\left(\frac{h\left[A^{(h-2)}(\theta) \right ]^{2}}{[A^{(h)}(\theta)]^{2}} -\frac{A^{(h-2)}(\theta)A^{(h-4)}(\theta)}{A^{(h)}(\theta)}2(h-2)\right.\right.\right. \\  \left.\left.\left.+(h-4)A^{(h-6)}(\theta)\right )\frac{\sin^{3}2\theta }{2}\right ]\frac{(\sqrt{6})^{h}}{2^{h}.8}e^{\frac{h\psi}{\sqrt{6}}}\left(1-he^{-\sqrt{\frac{2}{3}}\psi} \right )\dot{\theta}^{2}+\frac{\left[A^{(h-2)}(\theta) \right ]^{3}}{[A^{(h)}(\theta)]^{3}}\frac{3h^{3}}{16}\sin^{3}2\theta \dot{\theta^{2}}\right. \\  \left.-\left(\frac{\left[A^{(h-2)}(\theta) \right ]^{2}}{A^{(h)}(\theta)}-A^{(h-4)}(\theta) \right )\frac{(\sqrt{6})^{h-1}}{2^{h}.4}h\sin^{2}2\theta e^{\frac{h\psi}{\sqrt{6}}}\left(1+(2-h)e^{-\sqrt{\frac{2}{3}}\psi} \right )\dot{\psi}\dot{\theta}\right\}\simeq-V_{\theta}
	\end{multline}
	As from the slow-roll regime it is evident that the kinetic energies of the fields are much small compared to the potential, one can conclude that the term $\dot{\psi}^{2}$ in Eq.(\ref{approximated_psi_equation}) and the terms $\dot{\psi}\dot{\theta}$ and $\dot{\theta}^{2}$ in the equation Eq.(\ref{approximated_theta_equation}) are negligible compared to the other terms with respect to the constant coefficient. We also assume field accelerations $\ddot{\psi}$ and $\ddot{\theta}$ can be neglected with respect to the potential gradient. As a result the equation of motion for the fields takes the form which is shown in Equations Eq.(\ref{very_approximated_eq_motion for_psi_field}) and Eq.(\ref{very_approximated_eq_motion for_theta_field}).
\end{appendices}


\end{document}